\documentclass{emulateapj}

\usepackage{natbib}

\usepackage{graphicx}
\usepackage[space]{grffile}
\usepackage{latexsym}
\usepackage{amsfonts,amsmath,amssymb}
\usepackage{url}
\usepackage[utf8]{inputenc}
\usepackage{fancyref}
\usepackage{hyperref}
\hypersetup{colorlinks=false,pdfborder={0 0 0},}
\usepackage{color,soul}
\usepackage{verbatim}

\usepackage{longtable,array}

\usepackage[utf8]{inputenc}
\usepackage[T1]{fontenc}
\usepackage{lmodern}

\shorttitle{X-ray Spectral Analysis of GRS 1915+105}
\shortauthors{Peris {\it et al.}}

\begin{document}

\title{X-ray Spectral Analysis of the Steady States of GRS 1915+105}

\author{Charith S.\ Peris\altaffilmark{1,2},
	Ronald A.\ Remillard\altaffilmark{3},
	James F.\ Steiner\altaffilmark{1},
  	Saeqa D.\ Vrtilek\altaffilmark{1}, \\
	Peggy {Varni{\`e}re}\altaffilmark{4},
	Jerome {Rodriguez}\altaffilmark{5},
	Guy Pooley\altaffilmark{6}}
  
\altaffiltext{1}{Harvard-Smithsonian Center for Astrophysics, 60  Garden Street, Cambridge, MA 02138, U.S.A.; email:cperis@cfa.harvard.edu}
\altaffiltext{2}{Department of Physics, Northeastern University, 360 Huntington Avenue, Boston, MA 02115, U.S.A.}
\altaffiltext{3}{Kavli Institute for Astrophysics and Space Research, MIT, Cambridge, MA 02139, U.S.A.}
\altaffiltext{4}{APC, AstroParticule \& Cosmologie, UMR 7164 CNRS/N2P3, Universit{\'e} Paris Diderot, CEA/Irfu, Observatoire de Paris, Sorbonne Paris Cit{\'e}, 10 rue Alice Domon et L{\'e}onie Duquet, 75205 Paris Cedex 13, France}
\altaffiltext{5}{Laboratoire AIM, CEA/IRFU-CNRS/INSU-Universit{\'e} Paris Diderot, CEA DSM/IRFU/SAp, Centre de Saclay,
91191 Gif-sur-Yvette, France}
\altaffiltext{6}{Astrophysics, Cavendish Laboratory, Madingley Road, Cambridge CB3 0HE, UK}

\bibliographystyle{apj_hacked}

\newcommand{\hilight}[1]{\colorbox{yellow}{#1}}

\newcommand{\vdag}{(v)^\dagger}
\newcommand{\na}{New Astron.}

\newcommand{\msun}{{\rm M}_\sun}
\newcommand{\Msun}{{\rm M}_\sun}

\newcommand{\Rin}{$R_{\rm in}$}
\newcommand{\Efold}{$E_{\rm fold}$}
\newcommand{\Tmax}{$T_{\rm max}$}
\newcommand{\fcol}{$f_{\rm col}$}
\newcommand{\fsc}{$f_{\rm sc}$}
\newcommand{\Ldisk}{$L_{\rm disk}$}


\newcommand{\gallolikeslope}{$\eta \sim 0.68 \pm 0.35$}
\newcommand{\interceptgallo}{5.32}

\newcommand{\plateauslope}{$\eta \sim 1.12 \pm 0.13$}
\newcommand{\interceptplateau}{-10.28}

\newcommand{\hardradiogtfivemiliJy}{95 \%}
\newcommand{\hardradiogttwomiliJy}{99 \%}
\newcommand{\softradioltfivemiliJy}{99 \%}
\newcommand{\softradiolttwomiliJy}{77 \%}


\begin{abstract}
We report on the X-ray spectral behavior within the steady states of GRS 1915+105. Our work is based on the full data set on the source obtained using the Proportional Counter Array on the Rossi X-ray Timing Explorer and 15 GHz radio data obtained using the Ryle Telescope. 
The steady observations within the X-ray data set naturally separated into two regions in the color-color diagram and we refer to them as steady-soft and steady-hard. GRS 1915+105 displays significant curvature in the coronal component in both the soft and hard data within the {\it RXTE}/PCA bandpass. A majority of the steady-soft observations displays a roughly constant inner disk radius (\Rin), while the steady-hard observations display an evolving disk truncation which is correlated to the mass accretion rate through the disk. The disk flux and coronal flux are strongly correlated in steady-hard observations and very weakly correlated in the steady-soft observations. Within the steady-hard observations we observe two particular circumstances when there are correlations between the coronal X-ray flux and the radio flux with log slopes \gallolikeslope~and \plateauslope. They are consistent with the upper and lower tracks of \citet{Gallo12}, respectively. A comparison of model parameters to the state definitions show that almost all steady-soft observations match the criteria of either thermal or steep power law state, while a large portion of the steady-hard observations match the hard state criteria when the disk fraction constraint is neglected. 
\end{abstract}


\section{Introduction}

The primary source of information about stellar mass black holes in the Milky way and nearby galaxies is the radiation from mass accretion systems known as black-hole binaries (BHBs). They consist of a secondary donor star that transfers matter on to the black hole, creating an accretion disk which is X-ray luminous. Today we have $> 20$ binary systems which have a dynamically determined primary mass above 3 $M_{\odot}$: strong evidence to support the presence of a black hole. 

The X-ray spectra of most BHBs are readily describable using a simple model consisting of a multi-temperature accretion disk component together with a hard X-ray power-law component which is widely attributed to inverse-Compton scattering of disk photons in a hot corona and modeled using a simple power-law or cut-off power-law function \citep{McClintock06a}. In most cases a Fe K$\alpha$ emission line should be included at $\sim 6.4$ keV. Sometimes it is necessary to add a disk reflection component when the inclination angle is $\le$ 60~$^{\circ}$. In some cases absorption features are also evident (\citealt{Ueda98}, see also \citealt{Sobczak00}). Broadly speaking, this class of models has been employed to satisfactorily fit many spectra of numerous BHBs (e.g. LMC X-1, LMC X-3, GX 339-4, \citealt{Ebisawa91}; GS 2000+25, \citealt{Ebisawa91,Takizawa91,Terada02}; GS 1354-64, \citealt{Kitamoto90}; and Nova Muscae 91, \citealt{Ebisawa94}; GRO J1650-40, \citealt{Sobczak99a}; XTE J1550-564, \citealt{Sobczak00}) and have been an essential tool in forming a physical picture of these sources. Such endeavors at spectral fitting via such simple modeling has revealed that while individual BHBs have their own behavioral tendencies, BHBs generally occupy a few distinctive X-ray spectral-timing states (e.g., \citealt{Remillard06} hereafter RM06, \citealt{Fender04}).  

A black hole in outburst typically occupies one of three states: the thermal state, the hard state, or the steep power-law state (SPL; RM06). Loosely defined, the thermal state (also called the high/soft state) features domination of the X-ray spectrum by a hot accretion disk. The hard state (or low/hard state) features domination by a hard X-ray corona that is related to a steady radio jet. The SPL state (sometimes referred to as the very high state) shows a significant contribution from a steep power-law component which is linked to activity in a hot corona, the absence of a steady jet, but possible production of a transient jet and a prominent disk. These states have been pivotal in understanding the physics that generates different modes of black-hole accretion. 

Of the BHBs discovered thus far, GRS 1915+105 (hereafter referred to as GRS1915) stands out as exceptional in many ways. It was discovered in 1992 by the WATCH all-sky monitor on board GRANAT as a transient Galactic source \citep{Castro-Tirado92}, and it sparked a lot of interest as the first Galactic object discovered to exhibit superluminal motion in its radio jets \citep{Mirabel94,Fender99b}. These Galactic radio jet sources, which display bi-polar radio emission analogous to that observed in radio-loud active galactic nuclei, are known as ``microquasars''. Initial estimates suggested a $\sim 14 M_{\odot}$ black hole \citep{Greiner01}. More recently, a trigonometric parallax was measured for the radio nucleus yielding a distance of $8.6_{-1.6}^{+2.0}$ kpc and a revised primary mass of $12.4_{-1.8}^{+2.0}$ $M_{\odot}$ \citep{Reid14}. Using a quantitative definition of the thermal state, \citet{McClintock06b} selected 22 disk dominated observations and calculated its spin to be $a_{*} > 0.98$, establishing GRS1915 as an extreme Kerr-hole.

Among the fundamental characteristics that set this black hole apart from the rest is GRS1915's wild X-ray variability, the diversity of which has not been replicated in any other stellar-mass black hole. The complex X-ray light-curves of GRS1915 span at least 14 different variability classes \citep{Belloni00,Klein-Wolt02,Hannikainen05} and were thought to be unique until the recent discovery of a few similar patterns in IGR J17091-3624 \citep{Altamirano11}. However, even though extreme variability is commonplace in its light-curve, about half of the observations of GRS1915 show fairly steady X-ray intensity (see Section~\ref{sec:xraydata}), and most of these intervals yield spectra and power density spectra (PDS) that seem to resemble one of the three states (see Section~\ref{sec:statecomp}; see also \citealt{Muno99,Klein-Wolt02,McClintock06a}). This suggests that within the complexity of this source is a simpler underlying basis of states which may map to those observed in canonical BHBs.

Historically, X-ray spectral analyses of the hot inner accretion disks in thermal states have shown that as the flux varies by several orders of magnitude, the value of the inner disk radius (\Rin) remains remarkably constant \citep{Tanaka95}. This is also true for SPL state observations with low or modest Compton scattering fractions \citep{Steiner09b}. The stability of \Rin\ suggests a relationship to the innermost stable circular orbit ($R_{\rm ISCO}$), as prescribed by General Relativity. In the hard and quiescent states, however, direct measurements of the thermal emission from the inner disk have been elusive. The low temperatures of hard state disks makes them nearly undetectable for many instruments (e.g., XTE J1118+480, $\sim 0.024$ keV, \citealt{McClintock01}; Swift J1753.5-0127, $\sim 0.2$ keV, \citealt{Miller06a}, $\sim 0.28-0.37$ keV, \citealt{Miller07}; GX339-4, $\sim 0.24$ keV, \citealt{Belloni13}; $\sim 0.22$ keV, \citealt{Shidatsu11}; Cyg X-1, $\sim 0.2$ keV, \citealt{Makishima08} ). This has led to studies in the hard state being focused primarily on the power-law component of the spectrum using the observed flux as an approximation to the total coronal flux. In this context, GRS1915 presents a unique opportunity. It displays a consistently hot disk even in its harder states \citep{Muno99}. We will also exploit this high temperature disk to separate the disk and coronal contributions to the spectrum and explore their behavior within both soft and hard observations.

Correlations between the radio and X-ray fluxes of hard and quiescent state BHBs have been a source of interest as they indicate a likely physical relationship between the radio jet and the X-ray emitting corona (and possibly the disk as well). After \citet{Corbel03} first reported the correlation over a wide flux range for GX 339-4, \citet{Gallo03,Gallo06} found that the same relationship ($L_r \propto L_X^{\eta}$ where $\eta = 0.58\pm0.16$) held for a number of hard state black holes. \citet{Gallo12} revisited this inquiry and critically investigated a growing number of outliers to the X-ray--radio correlation and found evidence for a distinct second track (see also \citealt{Coriat11}). The first track of $\eta = 0.63\pm0.03$ corresponded to the prior track and a second track was revealed with log slope $\eta = 0.98\pm0.08$. These two tracks are thought to reflect different accretion regimes within the hard state. \citet{Gallo12} did not include GRS1915 in their analysis. 

The X-ray--radio correlation in GRS1915 was investigated by \citet{Rushton10} using All-Sky Monitor (ASM) and Ryle telescope data. They found close coupling between the mechanisms that produced X-rays and the radio within hard observations displaying a steady radio jet  (``radio plateau'' states; \citealt{Muno01}), which they fitted with an index of $\eta \sim 1.7 \pm 0.3$. This index, which showed GRS1915 to be distinct from other BHBs, suggested that the dominant mode of hard-state accretion in GRS1915 is efficient, unlike canonical BHB hard states. We will employ our complete {\it RXTE}/PCA data set of GRS1915 to revisit and explore the X-ray--radio correlation, while being able to clearly distinguish the coronal flux from the total flux that is conveyed by the ASM. 

The spectrum of GRS1915 has been notoriously difficult to describe using standard BHB models (see \citealt{Muno99,McClintock06b,Titarchuk09}). Therefore, our main goal is to fit a majority of our data, which consists of an ensemble of locally steady conditions of GRS1915 (described in detail in Section~\ref{sec:xraydata}) with the simplest possible phenomenological models. We will discuss the correlations between the model parameters which will include addressing simple questions such as: ``How does the inner radius of the accretion disk behave?'', ``Does the disk exhibit any connection to the corona?''. ``Do the corona and the disk show a connection to the radio jet?''. We will then attempt to explain the physics revealed by the parameter variation. 

In Section 2, we describe our observations. In Section 3, we present an overview of the spectral modeling process terminating with a description of the successful models which we use to analyze the data. We present the results of our spectral analysis in Section 4, followed by a comparison of our observations to canonical BHB states and to the variable states of GRS1915 \citep{Belloni00} in Section 5. We discuss the implications of our results in Section 6 and end with a summary of our core conclusions.

\section{Observations}

\subsection{X-ray Data}
\label{sec:xraydata}

{\it RXTE} consists of three instruments; the Proportional Counter Array (PCA), the High-Energy X-ray Timing Experiment (HEXTE) and the ASM. The {\it RXTE}/PCA consists of five co-aligned proportional counter units (PCUs), each with a collecting area of $\sim$1600 cm$^{2}$ and a xenon-filled detector volume with three layers of signal anodes. It is sensitive in the range of approximately $2-60$ keV with energy resolution $\approx 18$ \% at 6 keV (see \citealt{Jahoda06}). 

The spectral fitting was conducted on data obtained using all layers of proportional counter unit 2 (PCU-2) because it was operating during almost every observation. The PCA background subtraction was conducted using the pcabackest task of the HEASOFT/FTOOLS package.  We used the composite bright/faint source model of \citet{Markwardt12}. 

In this study, the elemental spectra consist of continuous data segments that occur between the common interruptions imposed by Earth occultations and the passage of the {\it RXTE} spacecraft through the South Atlantic Anomaly. We further imposed a segment break when the number of PCUs was changed or to exclude data when one of the PCUs was not operating properly.

During its operational period {\it RXTE}/PCA observed GRS1915 $\sim1800$ times. We filtered this data for continuous data segments greater than 400 s in both standard 1 and standard 2 data modes and found 2565 intervals exceeding 5.2 Ms, with a mean exposure of 2.1 ks. After imposing a further requirement that PCU-2 spectra are available with standard 2 data, then the total number of data segments available became 2563.

We selected four energy bands A (2.2-3.6 keV), B (3.6-5.0 keV), C (5.0-8.6 keV) and D (8.6-18.0 keV) and defined soft color - or hardness ratio - HR1 as the count rate in B divided by the count rate in A; and hard color (HR2) as D/C. The four PCA energy bands are used to compute a color-color diagram (CD) and a hardness-intensity diagram (HID), with a scheme to
normalize the count rates in each band, throughout the times of PCA
observations (1996-2012).  Such normalized CDs and HIDs are shown for
accreting black holes and neutron stars in \citet{Muno02}, RM06 and \citet{Lin07}.  However, there is
one significant difference in the normalization strategy used in the
present paper.  Previously, it was assumed that the X-ray spectrum of
the Crab Nebula is invariant, and the PCA count rates in each energy
band were normalized to ensure that the Crab CD and HID display
constant values, except for statistical fluctuations.  However, it has been
shown that the Crab Nebula actually varies at the level of several
percent over timescales of months \citep{Wilson-Hodge11}. Their
work utilized the observations of several instruments, including the
{\it RXTE}/PCA.  It can be concluded that the physical model for the PCA
Instrument, which is used to normalize the PCA response files, is
``stiffer'' than the Crab variations, i.e., the models parameters change
on longer timescales, allowing the Crab variations to be seen.

In the present investigation, the PCA response files serve as the
basis for the 4-band normalizations. We simulated Crab-like spectra
({\sc xspec: fakeit}) at 30 day intervals between 1996 January and 2012
January. The simulated spectra adopted the mean PCA Crab spectral
parameters, i.e. a power law with photon index ($\Gamma$) 2.11, normalization
constant 10.55, and interstellar absorption $N_H = 3.45 \times 10^{21} cm^{-2}$ .  For each
simulated PCU spectrum, we extracted the count rates within the four
intervals noted above. We then normalized each band to global
constants, using piece-wise linear fits during each PCU gain epoch.
The targeted normalization constants, per energy band, are the same
values used in \citet{Lin07}, i.e., 550, 550, 850, and 570 counts/s.
The normalization parameters, per PCU and per time interval, can then
be applied to any PCA data.

There is one caveat to this process that concerns the lowest energy
band (channel A).  When the Crab observations themselves are
normalized by the response files, via the Crab-simulated spectra
described above, the mean values for the normalized energy bands agree
with the target values, except for channel A.  There is a very
significant discrepancy between the real Crab spectra and the
simulated spectra at energies below $\sim3$ keV, where there are
substantial numbers of counts.  This discrepancy underlies the common
practice of PCA users to adopt lower limits for spectral fitting that
are typically in the range $\sim$2.5-3.0 keV.  We choose to compensate for
this problem by imposing an additional normalization constant per
PCU and per gain epoch for channel A (only) that brings the real Crab
count rates in line with the target value. The additional correction factors to the A Channel are typically in the range 5-15 \%,
depending on PCU and time interval.  Such values are similar to or less than the factors
that normalize different PCUs during a given time interval, or the factors that 
relate different time epochs for a given PCU (e.g., across the times of gain changes 
or the times when the propane layer was lost for two of the PCUs.)

This paper is focused on the results of spectral fits for the data segments that we infer to be quasi-steady. We have found that the best way to define quasi-steady conditions is to consider three measurement quantities: the source fractional variations (i.e., rms / mean) in 1 s bins, the same measurement with 16 s bins, and the rms variations in the soft color (HR1).  Each of these quantities allows us to easily identify the types of variable light curves characterized by high amplitude cycles that operate at timescales of minutes or longer (e.g., variability classes $\beta, \nu, \mu, \theta, \lambda, \kappa, \rho$; see \citealt{Belloni00}).  However, there are overlapping values of fractional variability (e.g., at 1 s) between light curve types that would appear to be quasi-steady (e.g., class $\chi$, which displays hard-state flickering widely treated as quasi-steady), versus others that display tracks in the CD that suggest temperature variations which cannot be averaged (e.g., weaker episodes of classes $\gamma$ or $\delta$).  We report, in advance, that none of the conclusions offered in this paper are changed if we impose additional restrictions that eliminate data selections when there are appearances of weak $\gamma$ or $\delta$ characteristics in the light curves. Such cases are subtle, and they would amount to 1-2 \% and 4-5 \% of the steady intervals, respectively.

\begin{figure*}
\begin{center}
\includegraphics[width=70 mm, angle=-90, trim=30 0 0 0 ]{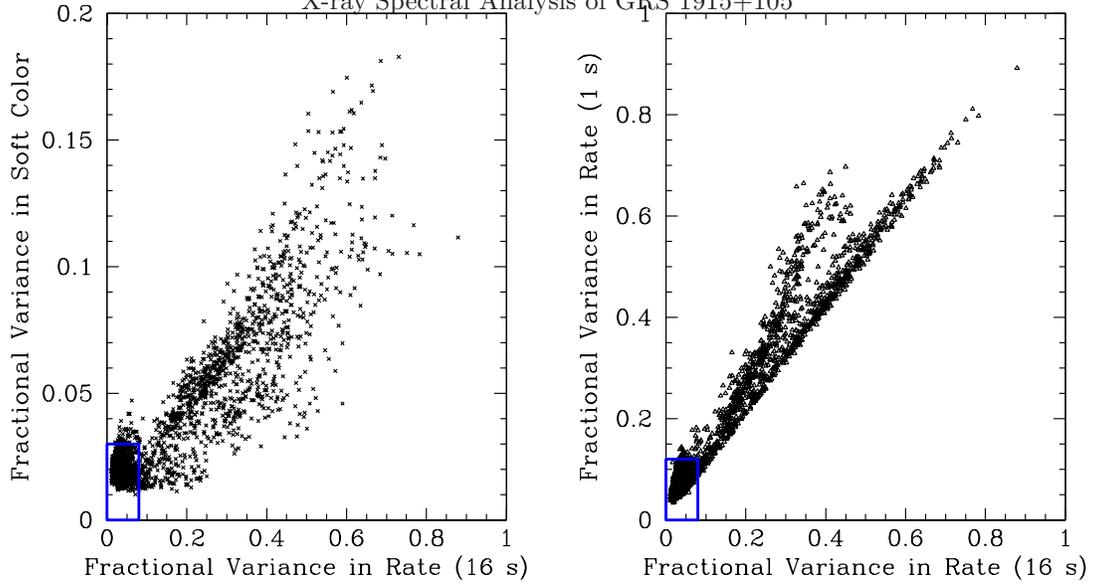}
\caption{Observations are termed ``quasi-steady'' if the fractional variability at 1 s is less than 0.12, the fractional variability at 16 s is less than 0.08, and the variations in HR1 (16 s) is less than 0.03. These observations fall within both blue boxes in the plots shown.}
\label{fig:var}
\end{center}
\end{figure*}

The values displayed by GRS1915 for the three variability quantities are shown in Figure~\ref{fig:var}.  We select segments as ``quasi-steady'' if the fractional variability at 1 s is less than 0.12, the fractional variability at 16 s is less than 0.08, and the variations in HR1 (16 s) is less than 0.03.  These selection criteria are displayed with blue lines in the lower-left corner of each panel in Figure~\ref{fig:var}.  There are 1257 data segments (which is 49\% of 2563 total) that lie in both selection boxes.  

\begin{figure*}
\begin{center}
\includegraphics[width=90 mm, trim= 60 0 0 0 ]{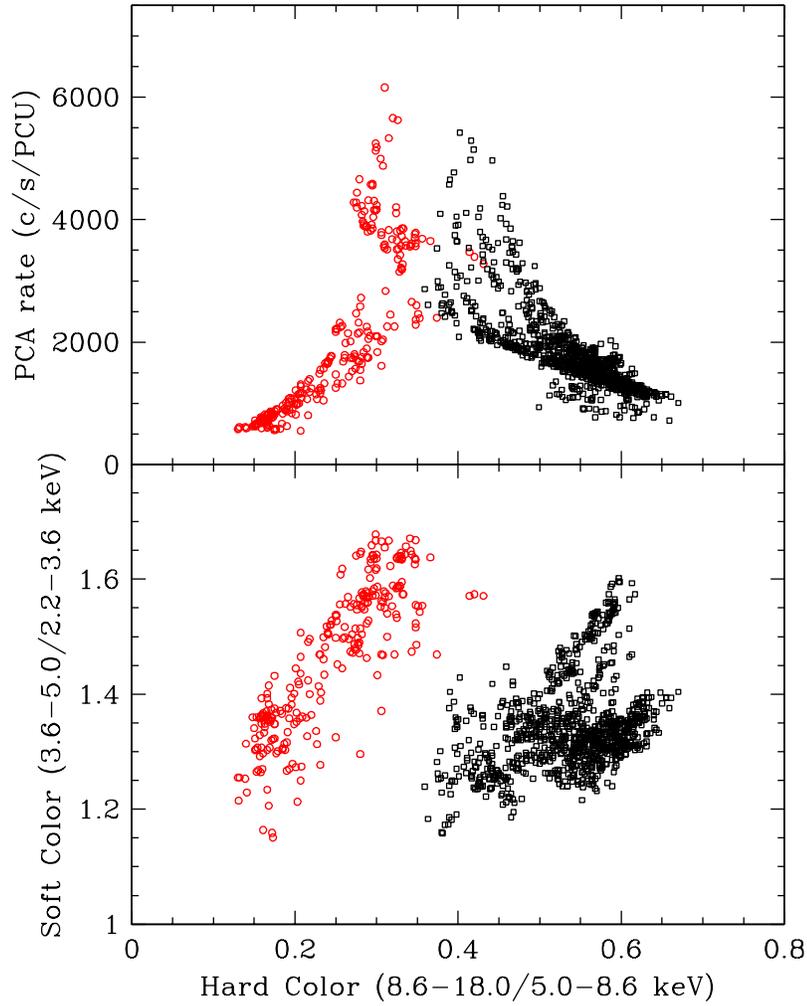}
\caption{The hardness-intensity diagram (HID; {\it top}) and the color-color diagram (CD; {\it bottom}) of the steady state data of GRS1915. The steady-soft observations are represented by red circles and the steady-hard observations by black squares.}
\label{fig:cc_ci}
\end{center}
\end{figure*}

The HID and CD of GRS1915 are displayed in Figure~\ref{fig:cc_ci} for the 1257 steady observations (2.55 Ms total exposure) that are the focus of this paper. The gaps between these steady observations show a broad distribution, ranging from 0.1 to 35 days (with an average of 2.2 days). We divide these observations into two groups which are best separated in the CD. We refer to the softer extended set of observations as ``steady-soft'' (red circles; 264 data points totaling 0.54 Ms) and the dense cluster of harder points as ``steady-hard'' (black squares; 993 data points totaling 2.01 Ms) throughout this paper. A tiny clump of points at [0.4, 1.55] in the CD which appeared ambiguous in the classification was identified as steady-soft, using the low integrated rms power (0.1-10 Hz) to discriminate. We note that this soft/hard separation is based only on the location of the observations in the CD and not on the more general thermal state and hard state classifications given by RM06. 

\subsection{Simultaneous Radio Data}
\label{sec:radioobs}

The Ryle telescope (now transformed into the AMI large array) located at the Mullard Radio Astronomy Observatory in UK, was an east-west radio interferometer primarily used for microwave-background studies. Its standard observations were 12 hours apiece so as to fill the synthesized aperture. In order to monitor sources with short-period variability, a program began using interstitial intervals between the standard (long) observations. This program, documented by \citet{Pooley97} collected nearly daily observations of GRS1915, at 15 GHz, from May 1995 to June 2006; providing a 10 year overlap with our {\it RXTE}/PCA X-ray observations. We used the data from this program to study the radio properties of GRS1915 corresponding to its behavior at X-ray wavelengths.

Although baselines between 18 m and 4.8 km were available in a variety of configurations, most of the data used in this study were obtained using a more compact array, typically less than 150 m. Observations were alternated with a nearby calibrator, B 1920+154, so that instrumental phase variations could be detected and removed. The flux-density scale was set using 3C 48 and 3C 286. The data were sampled every 8 seconds and averaged into five minute data points with an rms noise of $\sim$ 2 mJy \citep{Pooley97,Prat10,Rushton10}. We typically average over a longer time period in our study and our fractional uncertainty is reduced.

To investigate the X-ray--radio correlations, we explored either restrictively analyzing the strictly simultaneous observations or alternatively employing a larger sample for which the observation mid-times matched within $\pm0.5$ days. In each case, we took the mean value of all radio data points corresponding to each X-ray observation as the corresponding radio flux of that observation. Since the results in both cases were consistent, we employ the latter selection -- which is a significantly larger pool of data -- in our full analysis presented below. We found 70 and 624 radio-matched X-ray observations in steady-soft and steady-hard, respectively. 

\section{Spectral Modelling}
\label{sec:specmod}

In this section we describe the fitting process and the models employed for the spectral fits. All fitting was done using XSPEC 12.8.1 \citep{Arnaud96}. Spectra were fitted over the energy range 2.52 - 45 keV. The lower limit of 2.52 keV reflects the low-energy calibration limit of the detector, and the upper value of 45 keV is chosen due to calibration uncertainty and low effective area at higher energies. To improve the sensitivity of our fits we used a more recent and improved calibration tool {\sc \mbox{pcacorr}} \citep{Garcia14}. While {\sc \mbox{pcacorr}} comes with a recommended systematic uncertainty as low as 0.1 \% we employ the customarily used 1 \% systematic uncertainty (e.g. \citealt{Jahoda06}) for GRS1915 driven by simplicity of the models we employ.

After using the xspec fit command for the initial fitting, xspec's ``error'' command was used to find approximate confidence ranges for our main fit parameters. The error command varies the relevant parameter within its hard limits, and finds the change in value required to reach the $\Delta \chi^{2}$ which corresponds to the desired confidence interval for 1 d.o.f. and assuming Gaussian statistics. In some cases, as a byproduct of this detailed search for the confidence interval, a new minimum for the fit-statistic may be found.

Given the approximate and idealized nature of the models available to fit the spectra of BHBs, we delineated between successful ``good-fits'' and poor fits adopting a critical goodness-of-fit $\chi^{2}_{\nu} <$ 2 (see also \citealt{Steiner10,Steiner11}) which is appropriate for {\it RXTE}/PCA. We use only the ``good-fits'' for all analysis. The extremely large number of counts per spectrum in GRS1915 ($\sim$5 million) reduce channel errors significantly. This means that the uncertainty for many measurement bins is dominated by the systematic error, in which case our selection criteria admits errors (when using our simple models) of order $\sim$1.4\%.

\subsection{First model attempts}
\label{sec:modelattempts}

In this section we describe a subset of the models investigated to find the simplest ones that best fit the majority of the X-ray spectra. Up to 70 different model variants were explored with most producing unsatisfactory results. Typical BHB models incorporate a multi-temperature accretion disk coupled with a standard power-law component (a proxy for inverse Compton scattering) and a Fe-K$\alpha$ line component. We found that this simplest construction failed and that a strong requirement was the inclusion of a cutoff (i.e. curvature) with unusually low folding energy in the power-law component in order to produce good fits. This result established that the Compton component of GRS1915 (hereafter referred to in a more physical sense, as the `coronal component') is unusually curved within the PCA bandpass. An exponential cutoff was also used in analyses by \citet{Muno99} and \citet{Neilsen11}.

Gamma ray observations by \citet{Grove98} has detected emission from GRS1915 at energies up to $\sim 1$ MeV. X-ray observations by \citet{Zdziarski01} and \citet{Rodriguez08b} also confirmed the presence of a power-law like component at higher energies, while observing curvature in the spectrum consistent with our observations at energies $<45$ keV. Our fits indicate that an additional power-law added into our fits does not affect our $\chi^{2}_{\nu}$ values and number of good-fits. However, an added power law cannot be constrained given the limited bandwidth available using {\it RXTE}/PCA. Therefore we do not use it for our fits and do not venture to describe the spectrum of GRS1915 at energies higher than the PCA bandwidth. 
  
Next we explored a more physically-meaningful Comptonization model, {\sc comptt}, to fit the corona component \citep{Titarchuk94, Hua95, Titarchuk95}. In their paper modeling 107 spectra of GRS1915 in different spectral states, \citet{Titarchuk09} showed that photons from the disk that are scattered by the corona have a constant characteristic temperature $T_0 \sim 1$ keV. We found that by slightly adjusting their approach, so that the seed photon temperature was tied to the characteristic disk temperature $T_{\rm disk}$ (given by the component {\sc ezdiskbb}), we were able to produce a satisfactory number of good-fits. However, beyond this artificial scaling between the characteristic disk temperature and the seed photons, the model showed erratic behavior in $T_e$. In addition, both $T_e$ and the optical depth dropped below the limit where the analytical equations break down due to upscattering inefficiency (see figure 7 in \citet{Hua95}). For these reasons, we eschewed this model in favor of an empirical approach.

\subsection{The empirical approach to a power-law component: {\sc simplcut}}
\label{sec:simplc}

In order to empirically describe the curvature in the corona component within a self-consistent and flexible framework, we modified the {\sc simpl} model \citep{Steiner09}, to incorporate a high energy cutoff. {\sc simplcut} is an extension of the approximate Comptonization component {\sc simpl} with an additional parameter; a high energy exponential folding term -- \Efold. The parameter $\Gamma$ represents the asymptotic power-law before curvature comes into effect. This model is described by equations 3 and 4 in \citet{Steiner09} in tandem with an exponential folding in energy and will be described in detail in a future publication (Steiner et al., 2016). It was incorporated in xspec as a local model. This model has the virtue of being able to operate on an arbitrary seed spectrum, and conserves photon number -- as appropriate for Comptonization. In practice we tied the seed photons to the spectrum of the accretion disk. A fraction of the seed photons \fsc\ are scattered into a curved power law shape, locally computed for each energy. Importantly, this enables the calculation of an {\it intrinsic} disk luminosity, i.e., the luminosity emitted by the disk directly, prior to any transfer or scattering which  transpires in the corona. As already mentioned, this curvature is essential for achieving successful fits with GRS1915. {\sc simplcut}, like {\sc simpl}, also eliminates the unphysical rise at the lowest energies which is endemic in the standard {\sc powerlaw}  model. Our approach, though empirical and approximate, allows us to obtain self-consistent values for the disk-normalization parameter and thereby to make physical inferences about changes and patterns in \Rin .

\subsection{Steady-hard model}

As described in Section 2.1, the steady observations of GRS1915 naturally separate into two zones when represented in a CD or a HID (Figure~\ref{fig:cc_ci}). Due to the spectral differences in these two regimes, we fit them with slightly different models. The backbone of both models was {\sc simplcut $\otimes$ ezdiskbb}. 

In fitting the steady-hard data, the inclination of the system was fixed to $i=60^{\circ}$ \citep{Reid14} while the column density was fixed at $N_H = 5 \times 10^{22}$ cm$^{-2}$ (\citealt{Lee02} and references therein). We also tested $N_H$ values of 4.5 - 7.0 $\times 10^{22}$ cm$^{-2}$. No noticeable changes in the parameter distributions were evident and our conclusions are not sensitive to this setting.

The steady-hard data were modeled via\\
\\
{\sc tbabs $\times$ (simplcut $\otimes$ ezdiskbb + laor + gaussian)}.
\\
\\
The model is comprised of a number of elemental components: ({\sc tbabs}, \citealt{Wilms00}; {\sc ezdiskbb}, \citealt{Zimmerman05}; {\sc laor}, \citealt{Laor91}). Both {\sc laor} and {\sc gaussian} line centers were fixed at 6.5 keV. The Gaussian width is forced to be very narrow (width fixed to 10 eV), corresponding to a distant, nonrelativistic Fe reflection line, which is sometimes observed. The addition of this component results in a 10~\% increase in the number of good-fits. The {\sc laor} component describes reflection in the strong-gravity regime in which extreme relativistic effects cause a red and broad distortion of the line (examples of similar broad and narrow Fe-line emission have been commonly observed in AGN; \citealt{Brenneman14}). {\sc ezdiskbb} is a standard multicolor disk model, for which the inner disk edge has a zero-torque boundary. This model yielded an impressive $992/993$ good-fits ($99.9\%$). 

The yellow-filled boxes in Figure~\ref{fig:chidistribution} show the distribution of the goodness-of-fit for the steady-hard data. It is clear that despite the strong signal, spectral complexity, and our minimalistic models, we are able to successfully model an impressive majority of this data. 

\subsection{Steady-soft model}
\label{sec:ssfitting}

In fitting the steady-soft data, we employed the same inclination ($i = 60^{\circ}$), column density ($N_H = 5 \times 10^{22}$ cm$^{-2}$) and range of column density examined ($N_H = 4.5 - 7.0 \times 10^{22}$ cm$^{-2}$) as in the steady-hard observations, with no significant change in parameter distributions observed.

As noted in Section~\ref{sec:modelattempts}, we tested 70 model variants on GRS1915 spectra. This was driven by the need to add features, investigated one at a time, to gain acceptable fits for the steady-soft spectra. This is why the models for steady-soft and steady-hard observations differ. The extra terms have also been used for soft states, when required, by other researchers (see below).

The steady-soft data were modeled using\\
\\
{\sc tbabs $\times$ smedge $\times$ (simplcut $\otimes$ ezdiskbb + laor) $\times$ gabs}
\\
\\
They require the addition of a smeared absorption edge ({\sc smedge}; \citealt{Ebisawa94}, see also \citealt{Sobczak99a}), another feature of reflection, which onsets between 7.5-9.0 keV (the {\sc smedge} width was fixed to 7 keV). We additionally include a Gaussian Fe-absorption component ({\sc gabs}) with width fixed to 0.5 keV and central energy fitted between 6.3-7.5 keV. Fe-absorption in this region for the soft spectra of GRS1915 has been observed in previous studies \citep{McClintock06b} and is probably due to the presence of a strong disk wind \citep{Kotani00,Neilsen09}. $\chi^{2}_{\nu}$ was significantly improved by the addition of these components although the key model parameters (i.e. the disk's normalization and temperature) were not affected. 

Strong degeneracy within the model fits in the steady-soft data was revealed when attempting to determine parameter uncertainties via xspec's ``error'' command.  In particular, we found degenerate trends allowing in very high $\Gamma$ and high \fsc, as warned against as a modeling artifact in \citet{Steiner09}.  In order to proceed, and to mitigate this complication, we found it necessary to adopt a fixed value for $\Gamma$.  We investigated 3 values which were chosen based on typical $\Gamma$ values observed in canonical BHBs in soft states; 2.0, 2.5 and 3.0. While the parameter distributions remain relatively unaffected by the choice of value, we found that $\Gamma$~=~2.0 yields a lower number of good-fits while $\Gamma$~=~3.0 allows in observations which display degenerate tendencies. Hence we favor $\Gamma$~=~2.5. 

We also observed that a lower limit to \Efold\ of $\sim$10 keV is necessary for the success of this model. Although a significant number of our fits show \Efold\ cluster close to our lower limit, it serves to keep the coronal component from encroaching on the accretion disk and avoid degeneracy within our model fits.

\begin{figure}
\begin{center}
\includegraphics[width=1.0\columnwidth]{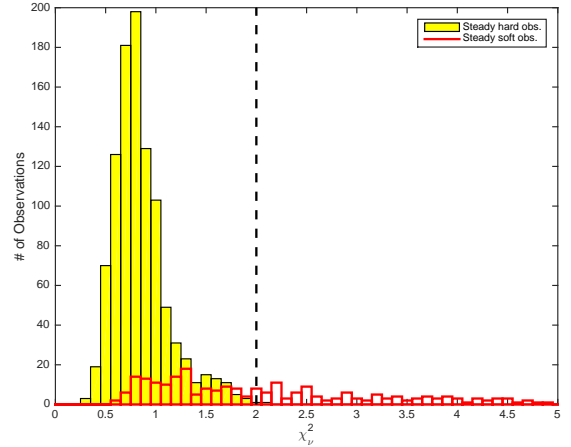}
\caption{The distributions of $\chi^{2}_{\nu}$ of steady-soft and steady-hard fits. The vertical dashed line at $\chi^{2}_{\nu} = 2$ represents the cut defined for a good-fit.} 
\label{fig:chidistribution}
\end{center}
\end{figure}

Using this model and set of constraints, we obtained $135/264$ good fits ($51\%$) for the steady-soft observations. The red-outlined boxes in Figure~\ref{fig:chidistribution} show the distribution of the goodness-of-fit for the steady-soft data. Admittedly, there is a broad distribution in $\chi^{2}_{\nu}$. We also acknowledge that the constraints we apply may limit the information we can obtain from our steady-soft model. However, due to the complexity of the steady-soft spectrum and the simplicity of the available models, we consider it {\it worthwhile} to examine the implied physical evolution of the steady-soft observations using our model. 

\begin{figure}
\begin{center}
\includegraphics[width=1.0\columnwidth, trim = 50 0 120 30]{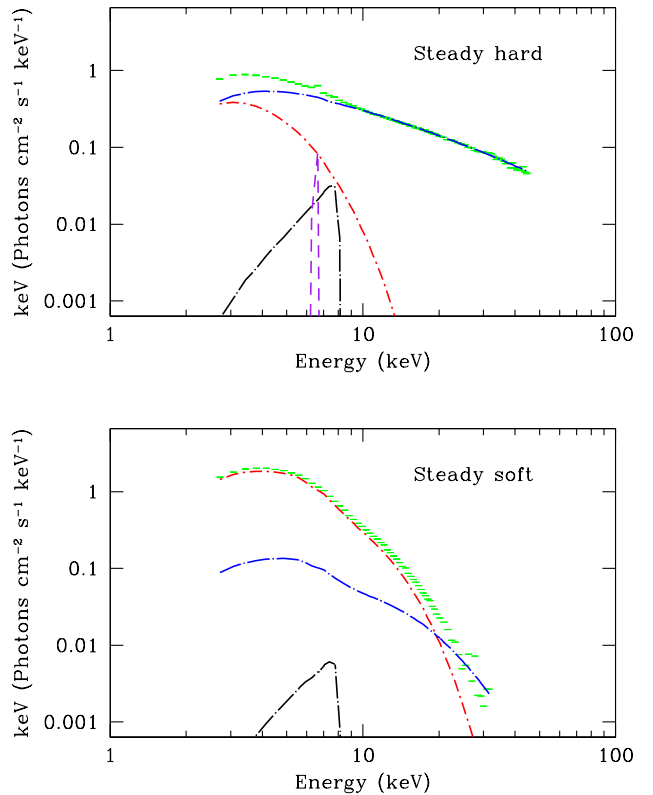}
\caption{Two unfolded spectra which are typical and exemplify the steady-hard and steady-soft data, respectively. {\bf Top:} The steady-hard spectrum for data segment 1103\textunderscore a (ObsID 20402-01-10-00) for which we obtain a fit with $\chi^{2}_{\nu} = 0.76$. {\bf Bottom:} The steady-soft spectrum for data segment 0881\textunderscore a (ObsID 10408-01-11-00) for which we obtain a fit with $\chi^{2}_{\nu} = 0.73$. Data is represented in green, the disk component in red and the coronal component in blue. Note the curvature in both coronal components.
}
\label{fig:spectra}
\end{center}
\end{figure}

Illustrative fits for both steady-hard (top panel) and steady-soft (bottom panel) are shown in Figure~\ref{fig:spectra}. Curvature is observed in the Compton components (see also \citealt{McClintock06a, Neilsen11}).

\section{Results}
\label{sec:results}

In this section, we present the results of our spectral analyses. All of the results described use only the good-fit observations characterized in Section~\ref{sec:specmod}. We report on the steady-soft and steady-hard behavior of the accretion disk, followed by the corona and then the radio jet. This is followed by a comparison of the observations of GRS1915 to canonical BHB states as well as to the variable states defined by \citet{Belloni00}. 

\subsection{The accretion disk}

The accretion disk parameters are of central interest to our investigation. Although the two models fitted to the steady-soft and steady-hard observations are different, each incorporates the multicolor disk model, {\sc ezdiskbb} \citep{Zimmerman05}, which has two free parameters: the maximum disk temperature ($T_{\rm max}$) and a normalization parameter ($K$). From the normalization, it is straightforward to determine \Rin. We emphasize that by using {\sc simplcut}, our model is self-consistent in counting Comptonized photons which enables us to directly obtain the {\it intrinsic} \Rin\ (in km) via:

\begin{equation}
R_{\rm in} = \left({\frac{K^{'} f_{\rm col}^4 D_{10}^2}{cos(i)}}\right)^{1/2}
\label{eq:rin}
\end{equation}

\noindent where {\it \fcol} is the color correction, {\it i} is the inclination and {$D$} is the distance to the source in kpc where $D_{10}=\frac{D}{10}$ (distance in 10s of kpc) and

\begin{equation}
K^{'} = \frac{K}{\left(1-f_{\rm sc}\right)}
\label{eq:norm}
\end{equation}

\noindent where \fsc\ is the scattering fraction determined by {\sc simplcut} (which is implemented so as to directly report $K'$). We note that many publications which do not use {\sc simpl} or its derivative models and thus have no means of conserving photon number use the apparent inner disk radius ($R_{\rm app}$) for their analysis. \Rin\ and $R_{\rm app}$ relate as:

\begin{equation}
R_{\rm in} = \frac{R_{\rm app}}{\left(1-f_{\rm sc}\right)^{1/2}}
\label{eq:rapp}
\end{equation} 
 
\noindent \Rin\ is calculated using a distance $D=8.6_{-1.6}^{+2.0}$ kpc, and inclination $i=60 \pm 5^{\circ}$ \citep{Reid14} and a constant color correction factor \fcol=1.7 \citep{Shimura95}. We caution that the measurements of \Rin\ are approximate and their accuracy is limited by the use of non-relativistic, classical models, and other simplifications including a single and static value for \fcol\ -- despite that it should vary with luminosity and depends upon mass and spin (e.g., \citealt{Davis06}). However, despite these uncertainties which may affect the scaling, our primary interest in the radius measurements is to test for large-scale variations, for which these higher-order modifications are moot.

\begin{figure*}
\begin{center}
\includegraphics[width=180 mm]{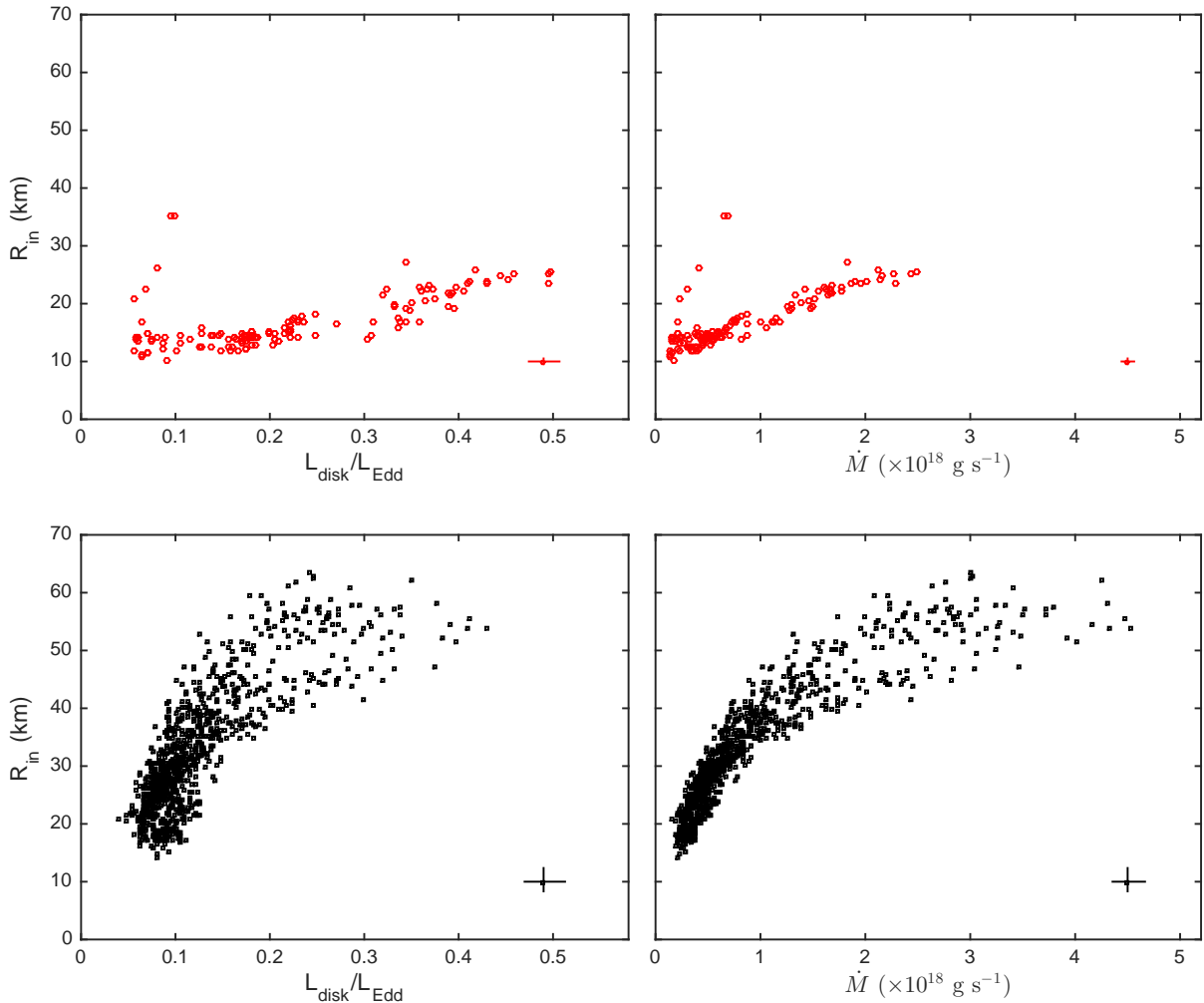} 
\caption{\Rin\ as a function of the disk luminosity (\Ldisk/$L_{\rm Edd}$ where $L_{\rm Edd} \approx 1.3 \times 10^{38} (M/M_{\sun})$ erg s$^{-1}$; \citealt{Frank92}) and mass accretion rate ($\dot{M}$). Top panels show steady-soft observations represented by red circles. A branch showing approximately constant \Rin\ is apparent. A few observations which show departure to larger \Rin\ are present at low luminosity. Bottom panels show the steady-hard observations represented by back squares. An organized increase in \Rin\ with increasing \Ldisk/$L_{\rm Edd}$ and $\dot{M}$ is observed. The mean 1~$\sigma$ error bars are indicated in the right-hand bottom corner of each plot. The full uncertainties (i.e. inclusive of the error due to distance and inclination) in the steady-soft and steady-hard parameters are as follows: $R_{\rm in} \sim 26$ \%,  $L_{\rm disk} \sim 55$ \% and $\dot{M} \sim 90$ \%.}
\label{fig:Rin}
\end{center}
\end{figure*}

\begin{figure*}
\begin{center}
\includegraphics[width=180 mm]{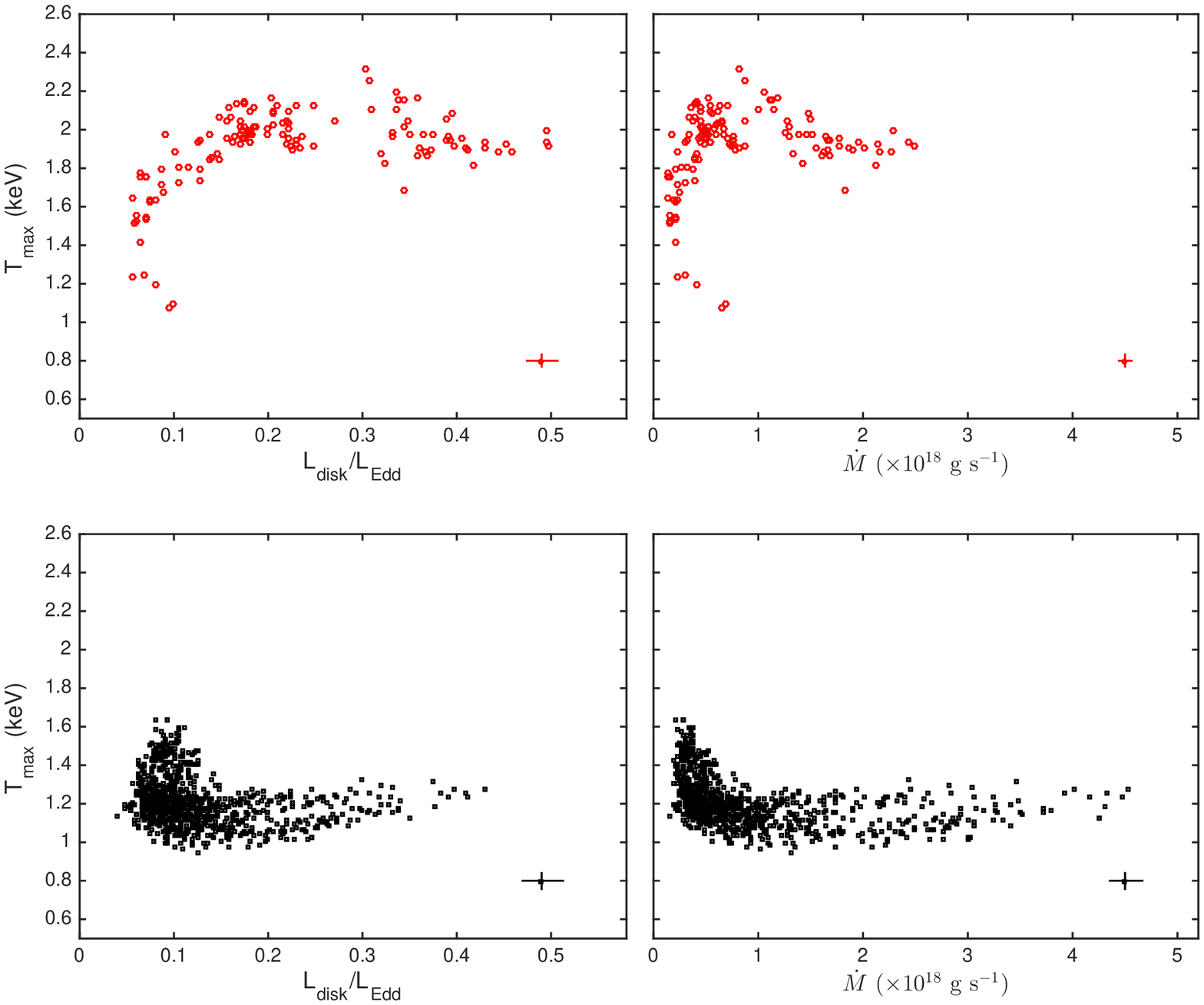} 
\caption{\Tmax\ as a function of \Ldisk/$L_{\rm Edd}$ and $\dot{M}$. Top and bottom panels show, respectively, steady-soft and steady-hard observations, as in Figure~\ref{fig:Rin}. Mean 1~$\sigma$ error bars are indicated as in Figure~\ref{fig:Rin}.}
\label{fig:T}
\end{center}
\end{figure*}

Figures~\ref{fig:Rin} \& \ref{fig:T} show the relationship between the disk temperature, the inner disk radius, the disk luminosity and the mass accretion rate. The mass accretion rate through the accretion disk is determined by

\begin{equation}
\dot{M} = {\frac{2 L_{\rm disk} R_{\rm in}} {G M}}
\label{eq:mdot}
\end{equation}

\noindent where

\begin{equation}
L_{\rm disk} = 73.9 \sigma \left( {\frac{T_{\rm max}}{f_{\rm col}}} \right)^4 R_{\rm in}^{2} 
\label{eq:ldisk}
\end{equation}

\noindent (see \citealt{Mitsuda84,Makishima86}). Red points (upper panels) represent the steady-soft observations while black points (lower panels) represent the steady-hard observations. The relationships between these accretion disk parameters obtained using both steady-soft and steady-hard observations show reasonably ordered branches. We remind the reader that these branches observed are not necessarily time-ordered. They simply represent a collection of steady state data. In real time, the source can disappear from these diagrams into variable states and later reappear at a different steady location. 

\subsubsection{The steady-soft accretion disk}
\label{sec:softdisk}

Figure~\ref{fig:Rin} displays a 4 cell grid of plots where we present the changes in \Rin\ as a function of two quantities: luminosity (expressed as a fraction of the Eddington limit) and mass accretion rate through the disk. The upper panels and lower panels show the steady-soft (red circles) and steady-hard (black squares) observations respectively.  Mean error bars for each plot are indicated in the right-hand bottom corner. Here and throughout, all errors presented describe 1 $\sigma$ confidence intervals.

A large majority of the steady-soft observations (96~\%) display a roughly constant \Rin, strongly reminiscent of the canonical BHB soft state (see also \citealt{Tanaka95}). Given the constancy of this branch over the large range of luminosities, its \Rin\ likely corresponds to $R_{\rm ISCO}$ for GRS1915. The slightly upward trend for \Rin\ at higher luminosities (upper panels in Figure~\ref{fig:Rin}) is reminiscent of the ``spin droop'' which has been observed in several BHBs including GRS1915 itself (\citealt{McClintock06b}; see also \citealt{Steiner10, Steiner11}). For this source, it is not likely to be caused by change in \fcol\ (see also \citealt{Done07}).  It could instead be the result of a ``local-Eddington effect'' \citep{Lin09} tied to the maximum and roughly constant inner disk temperature seen at \Ldisk\ $> 0.2 L_{\rm Edd}$ (see below). We cannot separate this effect from any structural changes in the disk that may occur at high \Ldisk.

At least six observations at low luminosity show a departure to larger \Rin\ compared to the constant \Rin\ branch\footnote{These are not the points at [0.4,1.55] in the CD which were close to the line between the hard/soft classification}. Amidst our quest for quelling degeneracy in the choice of steady-soft model (see Section~\ref{sec:ssfitting}), we investigated the possibility that these points may be erroneous. However, our analyses revealed these points to be just as robust as all the other points which exhibit a more canonical result. Therefore we proceed, albeit with caution, to treat them as good-fits and include them in our analysis in Section~\ref{sec:statecomp}. We also note that these points seem to lie in the \Ldisk--\Rin\ space that is occupied by the steady-hard observations. However, their classification is unambiguously steady soft, which is further shown in Section~\ref{sec:statecomp}.

Following the format of Figure~\ref{fig:Rin}, in Figure~\ref{fig:T} we present a second 4 cell grid of plots which displays the changes in \Tmax\ against the same two quantities as before, luminosity and mass accretion rate through the disk. The steady-soft observations (upper panels of Figure~\ref{fig:T}) show temperatures between 1-2.4 keV (consistent with those observed by \citealt{Muno99} for observations with no QPO; see their figure 8b) with the constant \Rin\ branch featuring higher temperatures of $\sim$2 keV. These temperatures are significantly higher than observed in other canonical BHB disks, for which a peak temperature of $\le1$ keV is typical (e.g. RM06). The observations at low luminosity which show departure to larger \Rin, also display the lowest temperatures.

\subsubsection{The steady-hard accretion disk}
\label{sec:harddisk}

The lower panels of Figure~\ref{fig:T} show the temperature of the inner disk for the steady-hard observations. They show a mean disk temperature of $1.19 \pm 0.12$ keV corroborating the results of earlier studies which suggest the presence of a hot disk even in the harder observations of GRS1915 \citep{Muno99}. This enables us to detect and measure the properties of the inner disk and explore its behavior. The most striking result (see Figure~\ref{fig:Rin}) is that \Rin\ for the {\it steady-hard} data increases with growing disk luminosity and mass accretion rate, demonstrating an evolving, truncated disk. 

Increases in \Rin\, at much shorter time-scales, have been observed in the variable states of GRS1915 ($\beta$ state, \citealt{Belloni97a,Belloni97b}; $\rho$ state, \citealt{Neilsen11}) suggesting a possible phenomenological link between the long ($>2.1$ ks) and short ($\sim 10$s of seconds) time scales in the system. This link will be further evaluated in Section~\ref{sec:statecomp}.

Several processes can be hypothesized as possible causes for this truncation and scaling:

\begin{itemize}
\item A local-Eddington effect for a thin disk
\end{itemize}
\begin{itemize}
\item An advection dominated accretion flow (ADAF)
\end{itemize}
\begin{itemize}
\item A magnetically-arrested disk (MAD)
\end{itemize}

A local Eddington effect has been invoked to interpret observations in
which the inner disk reaches a maximum temperature, and further
increases in luminosity are accommodated by increases in the disk
radius at constant (maximum) temperature.  Examples are several types
of behavior in the accreting neutron star subclass known as Z sources
\citep{Lin09} and the behavior of the disk in GRS1915
during the $\rho$ state variability cycles \citep{Neilsen11}. This concept was mentioned in the context of the gradient of the roughly constant \Rin\ branch of the steady-soft observations in Section~\ref{sec:softdisk}. However,
Figures~\ref{fig:Rin} and~\ref{fig:T} show that the steady-hard observations that show radius expansion exhibit temperatures 
a factor of two lower than the maximum temperatures observed in the system, and so we conclude that a
local Eddington effect is not a viable explanation for radius expansion in steady-hard conditions.

An ADAF is characterized by a low density, optically thin, quasi-spherical flow where most 
of the viscous energy released does not radiate efficiently, resulting in the energy being 
advected into the black hole. The ADAF model offers an alternative solution to
accretion in a thin disk, but the transition radius that may separate
these accretion geometries must be determined empirically, rather
than by deterministic factors (e.g., \citealt{Esin97}). Given the
organized behavior of the steady-hard truncated disk (Figure~\ref{fig:Rin}), it
would appear that the ADAF hypothesis, in and of itself, is incomplete. In
the absence of an ADAF transition mechanism that moves out with growing
disk luminosity, we do not discuss this alternative further.

The idea that global magnetic field properties may play a role
to distinguish steady-soft and steady-hard conditions
and magnetically truncate the accretion disk \citep{Narayan03,Tagger04} is plausible. The concept here is that a vertical or poloidal magnetic may be entrained into the inner disk region and then modify the geometry and energetics of the final stages of accretion. This is 
given further consideration in the Discussion (Section~\ref{sec:MAD}).

In Figure~\ref{fig:Rin} (lower panels), we note a possible bifurcation in the steady-hard data points when GRS1915 becomes luminous, $L\ge20$\% Eddington. The hint of split \Rin\ tracks maps to a similar appearance of splitting in the disk temperatures shown in Figure~\ref{fig:T}.  We know of no clear explanation for such a phenomenon, if it is real.

\begin{figure*}
\begin{center}
\includegraphics[width=180 mm]{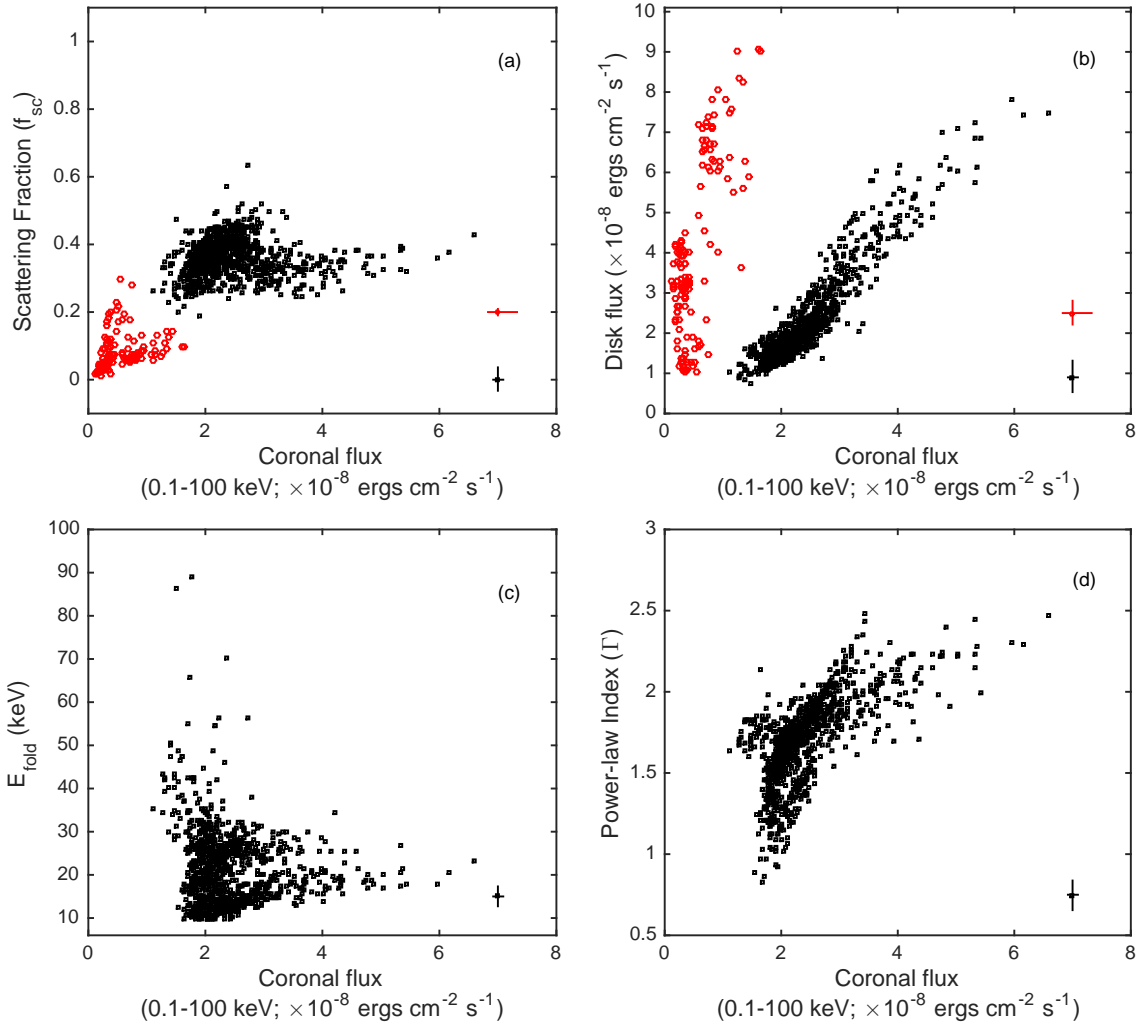}
\caption{Top panels show {\bf (a)} Scattering fraction (\fsc) and {\bf (b)} disk flux as a function of coronal flux in both steady-soft (red circles) and steady-hard (black squares) observations. Bottom panels show {\bf (c)} \Efold\ and {\bf (d)} $\Gamma$ as functions of coronal flux in the steady-hard observations.} 
\label{fig:corona}
\end{center}
\end{figure*}

\subsection{The disk--corona relationship}

While the origin of the soft X-ray component observed in BHB spectra has been generally agreed upon as thermal radiation from an accretion disk, the origin of the hard X-ray component (power law or cutoff power law) is still poorly understood. The prevalent paradigm has been inverse Compton scattering in a nebulous plasma referred to as the corona. In this scenario, the seed photons are from the inner disk but the origin and geometry of the corona is uncertain (e.g. \citealt{Poutanen99}). 

In Figure~\ref{fig:corona} (a,b) we explore the connection between the accretion disk and the corona. The use of the model {\sc simplcut} enables us to obtain the fraction of photons from the disk that contribute to the coronal flux (\fsc). This fraction is represented as a function of the coronal flux integrated over 0.1-100~keV. in Figure~\ref{fig:corona}a. \fsc\ stays low within the steady-soft observations. In the steady-hard observations \fsc\ is unusually well-behaved holding steady between $0.2-0.5$.  

Figure~\ref{fig:corona}b shows the division of spectral energy between the disk and coronal components. The steady-soft observations show comparatively low coronal flux and weak correlation between the disk and coronal components, with a slight growth in the contribution from the corona at high disk luminosity. While we fix $\Gamma$ at 2.5 for the fits represented here (see Section~\ref{sec:ssfitting}), we note that the coronal flux slightly decreases if a lower power-law index ($\Gamma$ = 2) is used. However, as stated in Section~\ref{sec:ssfitting}, the coronal flux distribution remains unchanged. In contrast, the steady-hard observations exhibit a more luminous corona and a strong correlation between the component fluxes (Figure~\ref{fig:corona}b; black points). The differences between these two tracks, despite considerable overlap in the flux from the accretion disk, could be taken to suggest that the coronae in the steady-soft and steady-hard conditions have different origins.

The strong correlation of the disk and coronal fluxes in the steady-hard observations suggests a connection between the processes that govern their production in GRS1915. Due to the extreme faintness of their disks, canonical BHBs in the hard state usually exhibit a near vertical track in the coronal flux verses disk-flux diagrams (see RM06). However, this correlation of disk and coronal flux might be a common feature of BHBs, which is undetected in canonical sources due to the faintness of their disks. X-ray observations of hard state BHBs with instruments sensitive to lower energies (e.g. NICER) will shed light on this matter in the future.

\subsection{The steady-hard corona}

\label{sec:corona}

Due to the constraints enforced on the steady-soft model (discussed in Section~\ref{sec:ssfitting}) we focus on the steady-hard observations, in order to explore the coronal parameters. The distributions of these parameters appear more complex and varied than those of the accretion disk parameters. 

In Figure~\ref{fig:corona} (c,d), we explore how the parameters related to the steady-hard coronal component, namely \Efold\ and $\Gamma$, vary as a function of coronal flux. Most of the steady-hard observations (80 \%) show a power-law index in the range $1.4<\Gamma<2.1$, which is one of the signatures of the canonical hard state (RM06). A more complete assessment of this is given in Section~\ref{sec:statecomp}.

We note the presence of a sub-population of points at low coronal flux that have significantly high \Efold\ values compared to the norm while also displaying well-constrained $\Gamma$ values ($\sim1.7$) with much less scatter. These observations display a move towards more typical hard state coronal characteristics (see Section~\ref{sec:x-ray-radio}). Although they occur at low flux values the high \Efold\ observations typically have $\sim$2.7 million counts per spectrum and their fits have been verified using the error command on xspec. The remaining observations display low \Efold\ values (between 10-30 keV), confirming the significant curvature in the Compton component within the PCA bandpass (Figure~\ref{fig:corona}c). \Efold\ shows no obvious correlation with the coronal flux. $\Gamma$ shows an overall increase with growing coronal flux (Figure~\ref{fig:corona}d).

Indications of the existence of two (or maybe more) coronal tracks within the steady-hard observations are observed within both \Efold\ and $\Gamma$ plots. We note the presence of two tracks at higher luminosities in Figure~\ref{fig:corona}c and two tracks with differing slopes in Figure~\ref{fig:corona}d. These tracks approximately match the set of tracks in Section~\ref{sec:harddisk}, but do not map one-to-one. Further interpretation of this splitting is left for future work.

\subsection{The radio jet}

In this section, we consider all radio observations that time-match (within $\pm0.5$ days) the steady-soft (70) and steady-hard (624) X-ray observations (see Section~\ref{sec:radioobs}).

\begin{figure}
\includegraphics[width=1.0\columnwidth]{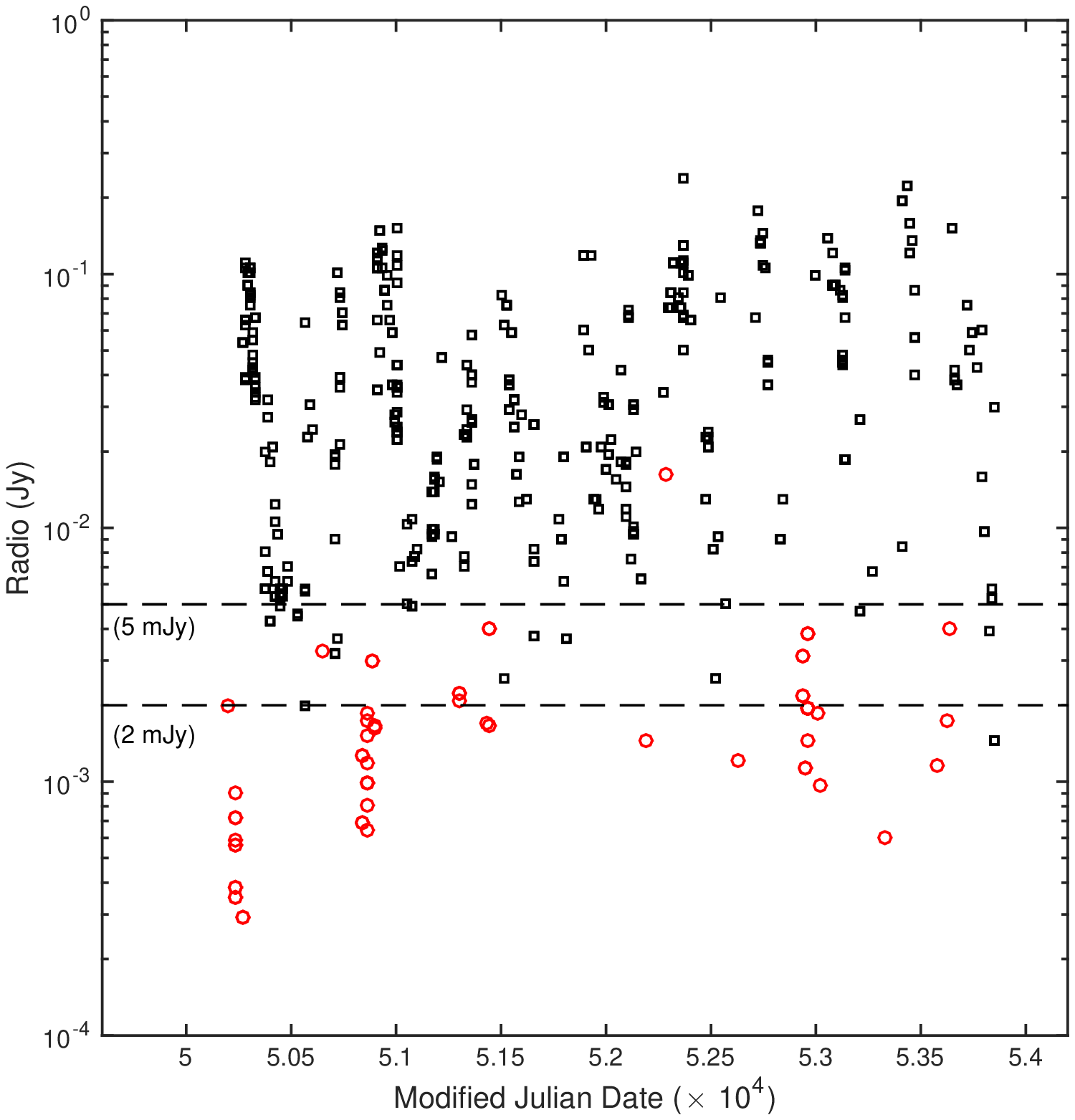}
\caption{The Ryle telescope's 15 GHz radio light curve of GRS1915's jet activity. Black squares show the radio flux corresponding with steady-hard observations and the red circles correspond to steady-soft observations (i.e. within $\pm 0.5$ days of the X-ray point). The dashed lines mark our reference flux levels of 5 mJy and 2 mJy. There is a clear difference in the brightness of the steady-hard compared to steady-soft data. We note that several steady-soft observations are not represented here due to extremely weak radio detections, or even non-detections. The single radio-bright steady-soft observation was tightly sandwiched between two hard states. Emission from jet material disconnected from the core possibly explains its higher radio flux. For reference, the typical error associated with each data point is $\sim \pm0.5$ mJy}
\label{fig:radlevels}
\end{figure}

We first examine how the radio flux levels differ in the steady-hard and the steady-soft observations. In Figure~\ref{fig:radlevels} we show the radio flux (represented in log scale) as a function of time. We find that \hardradiogtfivemiliJy~of the steady-hard data points have radio flux $> 5$ mJy with the maximum radio flux exceeding 100 mJy at times. \hardradiogttwomiliJy~of the steady-hard data points have radio flux $> 2$ mJy (consistent with \citealt{Muno01,Klein-Wolt02}). In the steady-soft data we conversely find that \softradiolttwomiliJy~have radio flux $< 2$ mJy and \softradioltfivemiliJy~have radio $< 5$ mJy. Our observations substantiate the idea that the steady-hard observations are associated with a radio jet. Some steady-soft observations show the presence of radio flux at lower levels. However, this low radio flux does not show any clear relationship with the steady-soft disk or corona. Furthermore, at low radio flux there is a chance that some portion of the emission arises from jet material which is detached from the core. This issue cannot be further investigated with these data given their insufficient spatial resolution. Hence we focus on the radio jet behavior of only the steady-hard observations.

\subsection{The corona--jet relationship}
\label{sec:corjet}

Strong coupling of the hard X-ray flux and the jet inferred from theory and observed in data (e.g. \citealt{Fender99a,Poutanen99}) supports the proposition that in hard states, the corona is associated with the base of the jet. The importance of this coupling is evident in the work by \citet{Gallo12} in uncovering two tracks relating X-ray and radio luminosity in multiple BHB systems. The study of this hard X-ray/radio connection may shed light on the process of jet formation. 

\begin{figure*}
\begin{center}
\includegraphics[width=120mm]{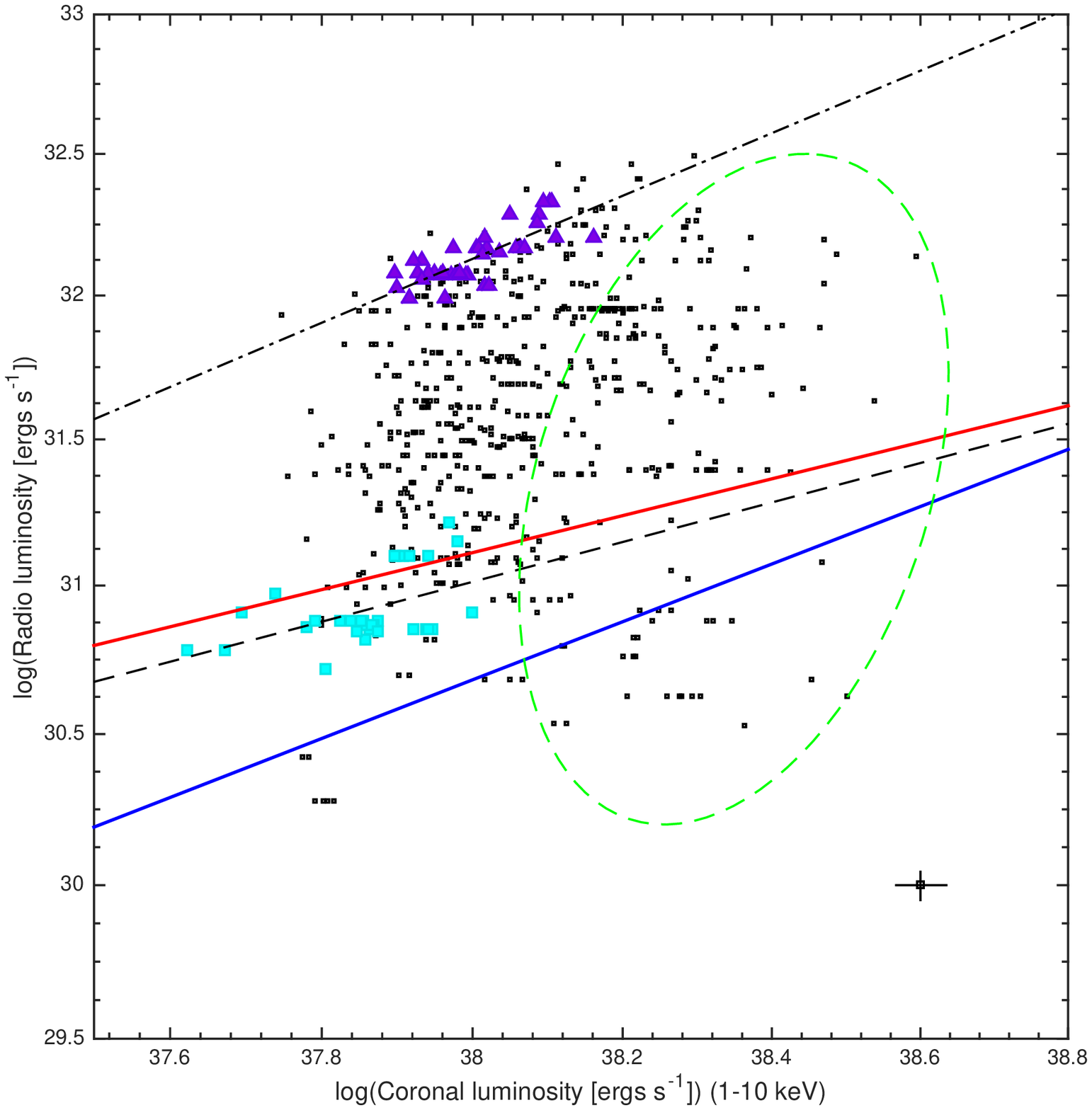}
\caption{Radio versus coronal luminosity. Cyan squares represent the G12-like points on the decline of the steady-hard light-curve dips observed within the time periods specified by Table~\ref{tab:Gallotimes}. Purple triangles represent the R10 points. The red and blue lines represent the G12~tracks ($L_r \propto L_X^{\eta}$ where $\eta$ = 0.63 and 0.98). The dashed line represents the fit to the cyan squares (\gallolikeslope) and the dot-dash line represents the fit to the purple triangles (\plateauslope). The region encircled by the dashed green line will be explained in Section~\ref{sec:diskjet}. }
\label{fig:logflux}
\end{center}
\end{figure*}

We isolate the coronal luminosity in the steady-hard observations of GRS1915 which produced good-fits (992/993), and investigate its behavior in response to the radio jet. Figure~\ref{fig:logflux} shows the radio luminosity as a function of 1-10 keV X-ray luminosity\footnote{To be consistent with the results of \citet{Gallo12} we use 1-10 keV integrated luminosities for this section.}. We also overplot the \citet{Gallo12} tracks (hereafter G12~tracks) in red (0.63 slope track) and blue (0.98 slope track). Taken as a whole, no obvious correlations are apparent. However, we identify three regions of interest: (1) the region with lowest coronal luminosity which seems to display a trend that matches the G12~tracks, (2) the region with highest radio luminosity which corresponds to a subset of the radio plateau state observations and (3) a diffuse region with high coronal luminosity which will be addressed in Section~\ref{sec:diskjet} (encircled by a green dashed line in Figures~\ref{fig:logflux} and~\ref{fig:rlcor}). In the following paragraphs we will discuss (1) and (2) in the context of recent literature. 

\begin{figure*}
\begin{center}
\includegraphics[width=180mm]{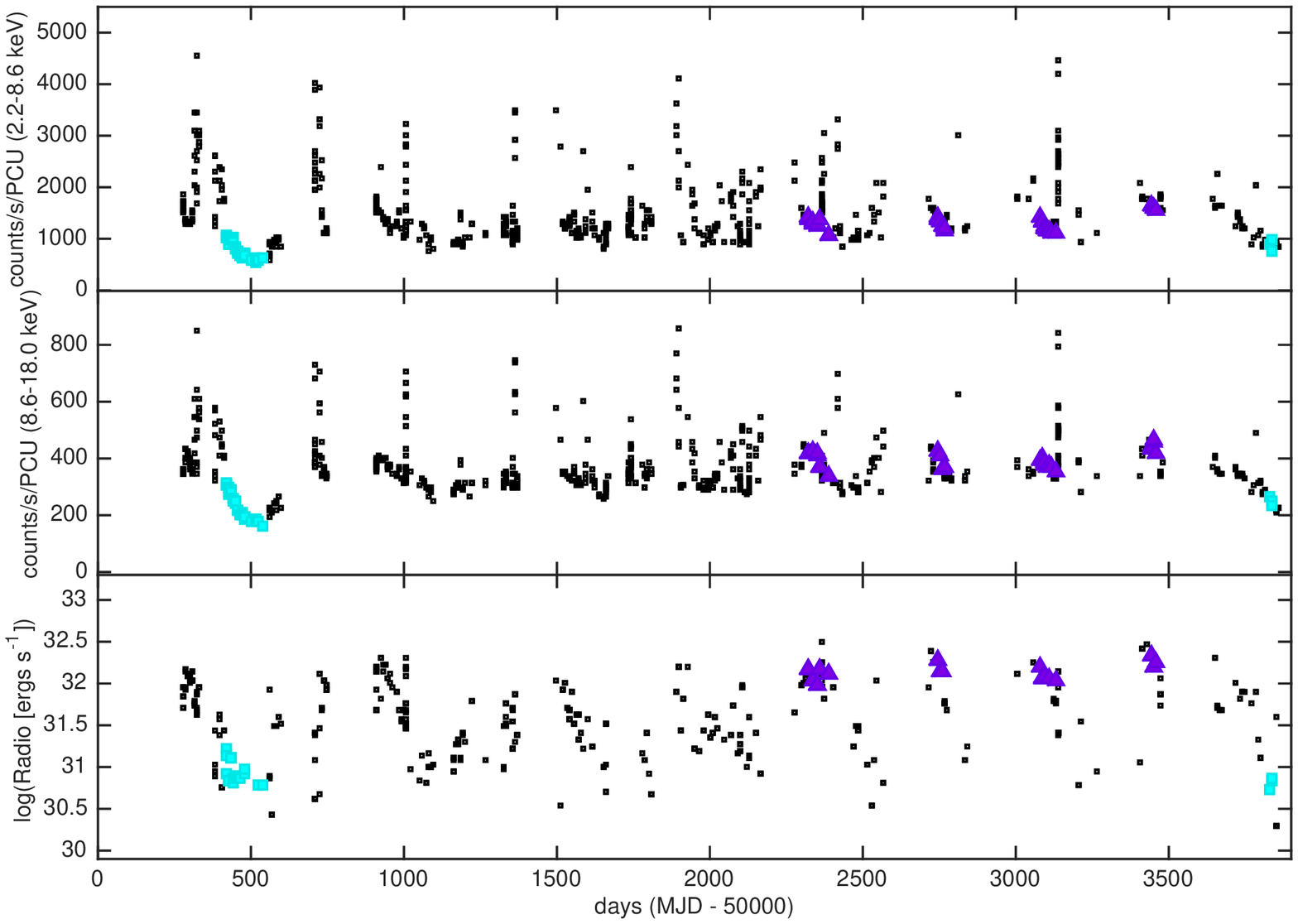}
\caption{X-ray and log(radio) light curves with G12-like points and the R10 points colored in cyan and purple respectively.}
\label{fig:logflux_lc}
\end{center}
\end{figure*}

The top two panels of Figure~\ref{fig:logflux_lc} show the steady-hard X-ray light curve in energy bands 2.2 - 8.6 keV and 8.6 - 18.0 keV. We note the presence of two time periods when the count-rate gradually decreases to low values in both bands. The first of these dips was categorized by \citet{Belloni00} as the $\chi_2$ class. The bottom panel in Figure~\ref{fig:logflux_lc} shows the radio light curve which indicates that the radio luminosity corresponding to the decline into these X-ray dips, is also relatively invariant. As the X-ray count-rate rises out of these light curve dips, the radio luminosity shows large variability ($\sim1$ order of magnitude). We neglect any such highly variable points. The points in the X-ray dips corresponding to the luminosity decline are isolated and shown in cyan in both Figures~\ref{fig:logflux} and~\ref{fig:logflux_lc}. They are defined by the time ranges given in Table~\ref{tab:Gallotimes}. Interestingly, we find that these points happen to lie almost exactly on the upper G12~track (red line in Figure~\ref{fig:logflux}). A fit to these points yields a log slope \gallolikeslope~(with the intercept at \interceptgallo\ where $L_X \propto L_R^{\eta}$). This is consistent with the slope of the upper G12~track, $\eta = 0.63$. However, we note that the uncertainty associated with our measurement is quite large. These points, hereafter referred to as ``G12-like'' points, also exhibit high \Efold\ values while showing $\Gamma$ values clustered tightly about $\sim1.8$, as typical for a canonical hard state. The G12-like points suggest that as the system declines to lower X-ray luminosities the steady-hard observations exhibit parameter values which are more typical of a canonical BHB hard state.

\begin{table}
\begin{center}
\caption{Time ranges used to obtain G12-like points}
\label{tab:Gallotimes}
\begin{tabular}{@{}lccc}
\hline
\hline
Start Date & End Date & \# of Obs.\\

\hline
50420 & 50550 & 27 \\
53820 & 53850 & 4\\
\hline
\end{tabular}
\end{center}
\end{table}

The second region of interest is the set of points which are a subset of the radio plateau states of GRS1915 which were previously selected for an independent X-ray--radio study by \citet{Rushton10}. We used {\it RXTE}/PCA observations with simultaneous radio data (see Section~\ref{sec:radioobs}) which are time matched within $\pm0.5$ days to the times of the their {\it RXTE}/ASM observations (\citealt{Rushton10}; private communication). Hereafter, we refer to them as ``R10'' points. They are represented by the purple triangles on the upper-half of Figure~\ref{fig:logflux}. A fit to the purple triangles in our plot yields a log slope \plateauslope~(with the intercept at \interceptplateau) showing a higher slope with respect to the G12-like points. 

In their study \citet{Rushton10}, obtained a log slope  $\eta \sim 1.7$. Their use of {\it RXTE}/ASM data necessarily limited their computation to an energy band of 2--12 keV and their analysis included both disk and coronal components. Our value is derived using the isolated coronal component luminosity (i.e. removing the contaminated disk contribution), integrated from 1-10 keV. It is therefore more suitable for comparison with the power law dominated BHB hard state luminosities used by \citet{Gallo12}.

\begin{figure}

\includegraphics[width=0.9\columnwidth]{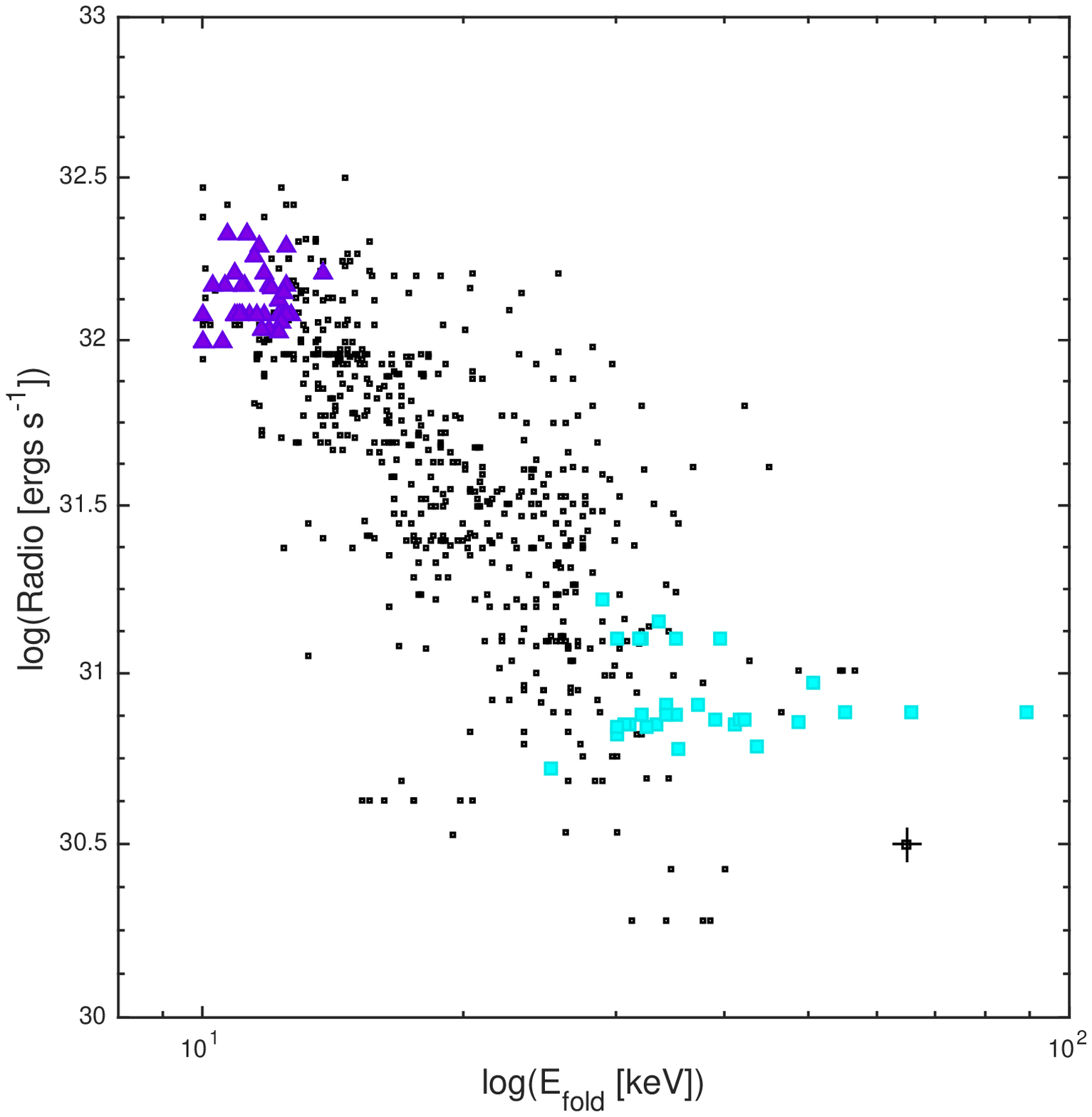}
\caption{\Efold\ as a function of radio luminosity. The colors represent the populations described in the caption of Figure~\ref{fig:logflux_lc}.
\\
\\}
\label{fig:lgradef}

\end{figure}

In Figure~\ref{fig:lgradef} we show the relationship of the radio luminosity and the model parameter \Efold. \Efold\ is the only parameter that displays a simple relationship to the radio luminosity, showing a strong non-linear anti-correlation. This anti-correlation nicely separates the G12-like points (cyan squares) and the R10 points (purple triangles), with the earlier having the highest \Efold\ values and the latter having the lowest \Efold\ values. The high \Efold\ values of the G12-like points, indicates a hotter corona, closer to a canonical hard state. The low \Efold\ values displayed by the R10 points and the anti-correlation itself, are still puzzles.

\subsection{A possible connection between the accretion disk and the radio jet}
\label{sec:diskjet}

\begin{figure}

\includegraphics[width=0.9\columnwidth]{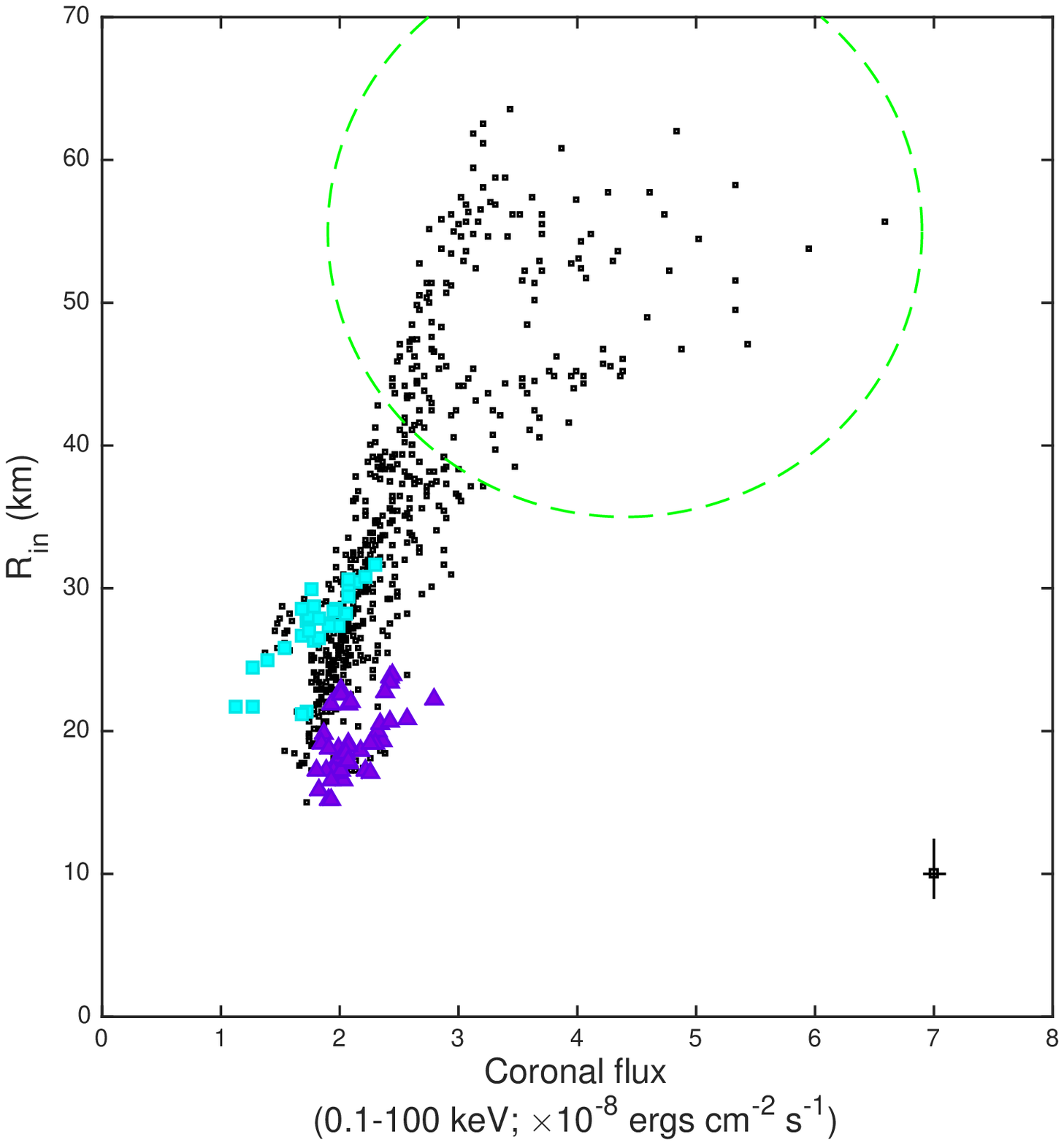}
\caption{The variation of \Rin\ as function of coronal flux. As in Figure~\ref{fig:lgradef}, the colored points represent the selections in Figure~\ref{fig:logflux_lc}. The dashed green-line roughly outlines the same points as in Figure~\ref{fig:logflux}. The apparent link of the two jet activity regions (cyan and purple points) to the two tracks does not hold in this region.}
\label{fig:rlcor}

\end{figure}

In Figure~\ref{fig:rlcor} we represent the two regions of different jet activity discussed above, in the context of the relationship between the coronal flux and the inner disk radius. It is interesting to note that both regions of jet activity that display correlations to coronal luminosity occurs when the steady-hard disk has low \Rin. They also appear linked to two tracks in Figure~\ref{fig:rlcor}. It is also noteworthy that, the black data points in Figure~\ref{fig:logflux}, situated in-between the G12-like region and the R10 region and outside the dashed green circle, displays mid-range \Efold\ values and forms a track in Figure~\ref{fig:rlcor} which exists in the middle of the purple and cyan tracks.

The apparent link of the two jet activity regions (cyan and purple points) to the two tracks does not hold at high \Rin. 
The extension of cyan and purple points in Figure~\ref{fig:rlcor} is very different from the extension of the same in Figure~\ref{fig:logflux}. While the cyan and purple points appear fairly organized in both Figures, any possible tracks within the green area in Figure~\ref{fig:rlcor} (high \Rin) are completely mixed when traced back to Figure~\ref{fig:logflux}. In summary, we note that while there appears to be a connection between the radio jet and the inner radius of the disk, the radio jet seems to be a function of other parameters which are presently unknown.

\section{Rigorous comparison to canonical BHB states}
\label{sec:statecomp}

In their review of BHBs, RM06 define three BHB spectral-timing states: thermal, SPL and hard. The criteria for each are given in their table 2 and are based on the characteristics of the X-ray spectra, power-density spectra (PDS) and quasi-periodic oscillations (QPOs). In this section we explore our observations of GRS1915 within their widely-used state framework.

\begin{figure*}
\includegraphics[width=190mm, trim=70 0 20 0]{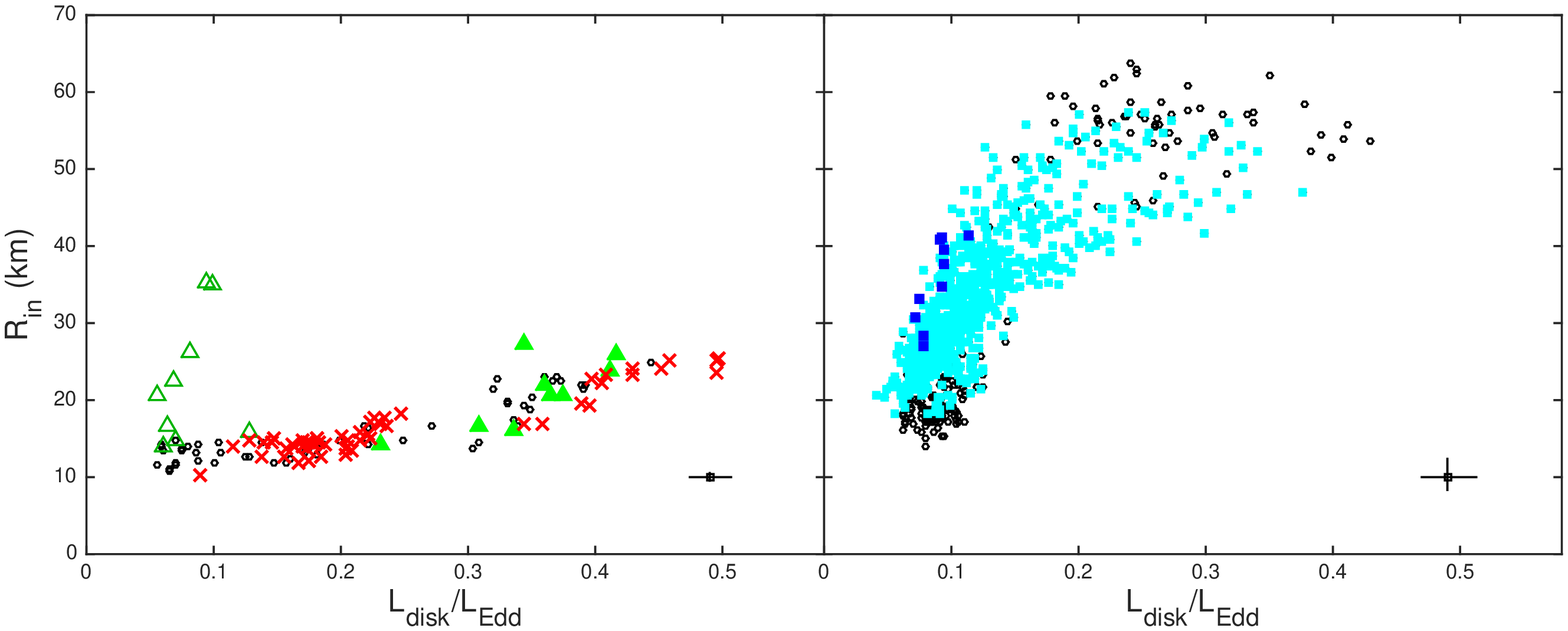}
\caption{{\bf Left:} Steady-soft observations with those matching the thermal (red crosses) and SPL state (green triangles) criteria marked. Filled-green triangles mark SPL state observations with LFQPOs while open-green triangles mark those with no LFQPOs. {\bf Right:} Steady-hard observations with those matching the hard (dark-blue filled squares) state criteria marked. Cyan squares show steady-hard data matching a relaxed definition of the hard state (see text for details). No steady-hard points matched the thermal or SPL state definitions and no steady-soft points matched the hard state definition. Black open circles mark observations which do not fit within the state classifications.} 
\label{fig:5br3st}
\end{figure*}

We took each X-ray observation and its corresponding PDS and matched their parameters to the criteria defining each state. Observations that matched the thermal criteria were represented as red crosses while observations that matched the SPL and hard criteria were represented as green triangles and dark blue squares respectively. The SPL state observations were split into two groups with those displaying low-frequency QPOs (LFQPOs; $< 30$ Hz) represented by filled-green triangles (referred to as SPL-qpo) and those without LFQPOs marked with open-green triangles (referred to as SPL-noqpo). Observations with parameters that did not fit within the boundaries of the state classifications were marked with black open circles. In addition, any steady-hard observations which did not fit the hard state disk fraction\footnote{Here, to be consistent with RM06, we use the disk fraction calculated using the apparent disk flux divided by the sum of the apparent disk and coronal flux, all in the 2 - 20 keV flux bands.} ($f_{\rm disk}$) constraint of $f_{\rm disk} < 20\%$ (see table 2 in RM06), but fit all other hard state criteria were represented as cyan squares. The resulting mapping of these data onto RM06 states is shown in Figure~\ref{fig:5br3st}, depicting \Rin\ verses disk luminosity.

A majority of the steady-soft data match either thermal (43 \%) or SPL (13 \%) states, while the remainder have one or more parameters with values that are outside of the criteria of the three states explored. The thermal states line up nicely, dominating the constant \Rin\ branch. This is consistent with the observations of canonical BHB thermal states which show \Rin\ going down to $R_{\rm ISCO}$ (e.g. \citealt{Steiner09b}).

Among the steady-hard observations, we find a few (0.01 \%) that are consistent with the hard state. However, we find that when the disk fraction is neglected, a majority of the steady-hard observations (80 \%) are found to match the hard state criteria (cyan points in Figure~\ref{fig:5br3st}). The higher disk fraction in GRS1915 is a consequence of the unique presence of a hot and bright disk in the steady-hard observations, as determined from our spectral fits. The alignment of the other parameters (rms power, spectral index and QPO behavior) with the well-studied states in many other black-hole systems confirms the posited association between the steady observations of GRS1915 and the canonical black hole states.

When compared, the two groups of SPL observations display several distinct differences in characteristics in addition to LFQPO presence. SPL-noqpo includes the observations that show departure in \Rin\ (see Section~\ref{sec:softdisk}). It also materializes at lower luminosities compared to SPL-qpo as seen clearly in Figure~\ref{fig:5br3st}. Furthermore, SPL-noqpo displays lower integrated rms power (0.1-10 Hz) and higher scattering fraction (\fsc) when compared to SPL-qpo.

\begin{figure}
\begin{center}
\includegraphics[width=1.0\columnwidth]{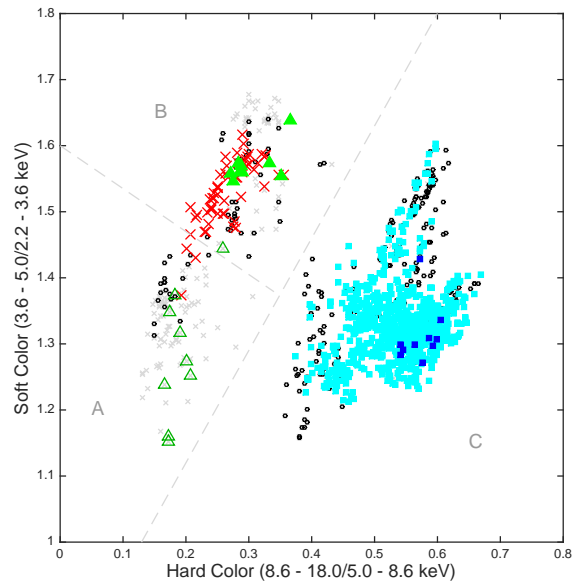}
\caption{The normalized CD, with symbols as defined in Figure~\ref{fig:5br3st}. Light-grey dashed lines roughly demarcate the regions ``A, B and C'' regions that govern fast variability cycles in GRS1915 (see also figure 8 in \citealt{Belloni00}). State assignments appear to map to the A, B and C regions. Light-grey crosses indicate observations that did not satisfy the goodness-of-fit criterion.}
\label{fig:comp_Belloni}
\end{center}
\end{figure}

Figure~\ref{fig:comp_Belloni} shows the normalized CD with data points color-coded for states, as described above (see Figure~\ref{fig:cc_ci}). The thermal and the SPL-qpo observations cluster together displaying higher HR1 compared to the SPL-noqpo observations. Remarkably, the locations of the two resultant CD clusters appear to be similar to the those of the A and B regions defined for GRS1915's {\it variable} observations by \citet{Belloni00} (see their figure 8). While the cluster comprised of the thermal and SPL-qpo observations maps to B, the SPL-noqpo observations appear in a similar location to A.  The steady-hard observations map to the location of the C variable state. If such a link exists between the steady observations and the transient A, B and C states, it demonstrates a compelling link between short and long time scales in GRS1915's phenomenology.

\section{Discussion}
\label{sec:disc}

In this section we discuss the confidence in our models and the implications of our results. We first establish our confidence in the models used in this paper by exploring several aspects of their performance. We then move on to discuss the results which they have enabled us to obtain. First we discuss the behavior of the accretion disk in the steady-soft observations. Next, we discuss a magnetically arrested disk as a possible scenario to explain the disk truncation observed in the steady-hard observations. We then discuss the X-ray--radio correlations we observe followed by possible effects due to high spin. 

\subsection{Model confidence}

Our adopted spectral models combine the paradigms of BHB accretion research with pragmatic adjustments to gain acceptable model fits. Since the models are not unique, we must evaluate the results on performance issues, e.g. the ability to provide physical parameters that are self-consistent, robust in handling the large range in source luminosity, productive in gaining insights into source behavior and capable of showing correlations with measurements independent of the spectral fits.

As shown in Section~\ref{sec:simplc}, we used two slightly different spectral models to fit the steady-soft and steady-hard observations. In the steady-hard observations we observe that \Rin\ drops to $\sim15$ km when closest to the black hole. This is consistent with the \Rin\ observed in the steady-soft data showing a self-consistent picture of \Rin\ across both models.

Our observation of a roughly constant \Rin\ branch in the steady-soft observations of GRS1915 over a wide range of luminosity, strengthens our confidence in the robustness of our model in handling changes in source luminosity. The minority of steady-soft observations that deviate from the main branch and show increasing \Rin\ have common properties; they are in the SPL state on the basis of a low disk fraction in energy flux, rather than the alternative path that is based on the presence of LFQPOs.

In order to explore parameter correlations with measurements independent of the fits we turn to timing information within the steady-hard observations. 100~\% of the steady-hard power density spectra display a strong LFQPO ($\sim 0.1-10$ Hz). We separate them into two groups based on the shape and maximum frequency of the LFQPO. Type 1 consists of single peak low amplitude higher frequency LFQPOs with no harmonics and Type 2 characterizes high amplitude lower frequency LFQPOs accompanied by one or two harmonics. 

\begin{figure}
\begin{center}
\includegraphics[width=1.0\columnwidth]{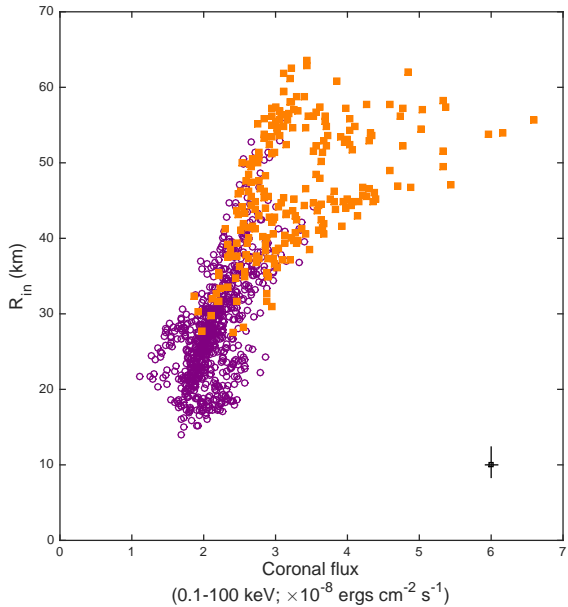}
\caption{Observations with Type 1 (orange squares) and Type 2 (purple circles) LFQPOs represented in the plot of \Rin\ verses coronal flux. The types are defined within the text.}
\label{fig:qpos}
\end{center}
\end{figure}

In Figure~\ref{fig:qpos} we represent observations corresponding to Type 1 and Type 2 LFQPOs colored in orange and purple respectively, within a plot showing \Rin\ as a function of coronal flux. The Type 1 and Type 2 LFQPO observations separate to low and high radii respectively. This shows that the timing information is tied to the structure of the inner disk as inferred from spectral results. The correlation of the model-dependent \Rin\ with independent LFQPO properties bolsters our confidence in our steady-hard model.

We also explored the robustness of the main results (i.e. an approximately constant \Rin\ branch in the steady-soft observations and a disk truncation in the steady-hard observations) by investigating parameter distribution in other tested models which fell short of our expectations and were not used in our analysis. We derived and compared the apparent \Rin\ and \Ldisk\ of our {\sc simplcut~$\otimes$~ezdiskbb} models with models that used components such as {\sc powerlaw~$\otimes$~highecut} and {\sc comptt}. The \Rin\ distributions produced by these models were found to reflect our main results for both steady-soft and steady-hard data sets. Such robustness of results, even though these models did not return sufficient good-fits, strengthen our confidence in our {\sc simplcut~$\otimes$~ezdiskbb} models.

\subsection{Behavior of the accretion disk in the steady-soft observations}
\label{sec:softdiskdiscussion}

\citet{Tanaka95} observed for three systems, that \Rin\ is remarkably constant over a wide range of bolometric disk flux. They associated this constant \Rin\ with the radius of the innermost stable circular orbit ($R_{\rm ISCO}$). An alternative way of displaying the same result has been to plot disk luminosity verses temperature which approximately goes as \Ldisk\ $\propto T^4$ for constant \Rin\ \citep{Gierlinski04,Kubota04,Done07}. Since then, the stability of \Rin\ when the system displays a soft state has been observed for many BHB systems and is regarded as a common trait of the soft thermal dominated state of BHBs. The validity of the claim of this stable \Rin\ as $R_{\rm ISCO}$ is bolstered by the many recent and successful spin measurements that are based on this assumption \citep{Shafee06, McClintock06b, Kolehmainen10, Steiner11, Steiner12, Steiner14}.
 
In Figure~\ref{fig:Rin}, we observe a similar trait displayed by 96 \% of the observations. It serves as a reminder that along with its intense variability, GRS1915 also displays properties similar to canonical BHBs. 

Within this roughly constant \Rin\ branch we note a slight increase in \Rin\ when luminosity rises above 0.3 \Ldisk/$L_{\rm Edd}$. This increase is manifested in the ``spin-droop'' found in several black hole systems including GRS1915 itself (e.g., \citealt{McClintock06b}; \citealt{Steiner10, Steiner11, McClintock14}), and may mark the onset of slim-disk effects (\citealt{Sadowski09, Sadowski11}; though see \citealt{Straub11}). The increase in \Rin\ at high luminosities results in a decrease in the inferred spin value which was apparent in the inconsistency of the two spin measurements for GRS1915 by \citet{McClintock06b} and \citet{Middleton06} (see \citealt{McClintock06b} for details). Considering that the approximately constant \Rin\ branch also displays nearly constant temperature which also is the highest temperature observed within the steady observations (Figure~\ref{fig:T}) the increase in \Rin\ (and corresponding ``droop'' in spin) might also alternatively be a result of a ``local Eddington'' effect much like that of the flaring branch of Z-sources \citep{Lin09}.

\subsection{A possible magnetic-field controlled truncation in the steady-hard observations}
\label{sec:MAD}

In this section we further discuss the idea presented in Section~\ref{sec:harddisk}, namely that global magnetic field properties may be the mechanism distinguishing steady-soft and steady-hard data; and that the magnetic state of the disk may account for its truncation when steady-hard.

Both theory and 3D Magnetohydrodynamical (MHD) simulations have predicted that the accumulation of strong poloidal magnetic field close to the black hole can disrupt the accretion of gas through the disk.  As poloidal field builds, magnetic pressure can at some point balance the ram pressure of the accreting gas, halting the flow and producing a truncated inner disk. Originally proposed by \citet{BisnovatyiKogan74}, this theory was refined by \citet{Narayan03} and the effect in question referred to as a Magnetically Arrested Disk (MAD). 

The presence of a global magnetic field was also invoked to explain the presence of X-ray QPOs in BHBs \citep{Tagger99, Varniere12}.  Here, magnetized disks were investigated in a theoretical framework known as the Accretion-Ejection Instability (AEI).  The degree of disk magnetization was later claimed to be the basis for the different X-ray states in BHBs \citep{Tagger04}, including the ``A, B, and C'' states that form the basis for fast variability cycles in GRS1915 (see end of Section 5 and also \citealt{Varniere11}). 

When accretion close to a black hole is disrupted by a MAD, it is suggested that matter flows inward in discrete blobs which travel much slower than the free-fall velocity. During this phase almost a significant fraction of the rest mass energy of the gas is expected to be released as heat, radiation and mechanical energy. Simulations have recently shown that MADs are also capable of producing jets efficient at extracting spin energy ($> 100 \%$ efficiency in ideal circumstances which in part requires $a_{*} > 0.9$, \citealt{Tchekhovskoy11, Mckinney12}). 

We note that thus far simulations of MADs have mostly treated accretion disks of substantial thickness, since a thick flow facilitates the trapping of magnetic field lines. They have also been concentrated on systems with low radiative energy loss. More study is needed to assess the importance of MADs in the thin-disk high-spin regime with high radiative losses. We remind the reader that GRS1915 is a system which belongs to the latter regime and therefore our suggestions should be taken with due caution.

Our results show that \Rin\ increases systematically with increasing disk luminosity and hence the estimated mass accretion rate through the disk. However, if the sampling of time-disparate points at a given disk luminosity suggest the same \Rin, then it would be reasonable to presume that the MAD must reach an equilibrium at a given $\dot{M}$, i.e., that B-field losses (reconnection and outflows) must balance the continual inflow of field lines seeded by the global field, so as to hold \Rin\ steady. In the MAD scenario, the \Rin--$\dot{M}$ correlation further suggests that the global B-field is not random, but is also regulated in some way by accretion, e.g., arising from some type of dynamo in the outer disk.  If not, changes in the global field strength would likely destroy the correlation, as we sample dozens of re-formed hard-steady conditions during the 16 years of {\it RXTE} observations. Additional questions concern the details of magnetic flux accumulation and the effects of disk thickness. Further work is needed to consider these issues and properly assess the MAD as an explanation for hard-steady conditions.


\citet{Varniere11} present another way to classify observations based on the kind of instability that occurs in the disk. This was first developed as a continuation of the extended Magnetic Flood Scenario \citep{Tagger04,Varniere07} which has been proposed to explain the overall behavior of the source based on the stored magnetic flux in the inner region of the system. 
The classification is based on two key parameters; the degree of magnetization of the disk ($\beta$) and the position of the inner edge of the disk ($\xi_{\rm int}$). From these two parameters they obtain four possible ``states''  harboring up to two simultaneous instabilities and having quite distinctive timing specificities. Their spectral differences have not been explored in detail but an overall behavior can be deduced from the different instabilities.   

The softer states are characterized with a lower magnetization level in the disk, while the harder states tend to have a fully magnetized disk. This clearly separates the behavior of both states even though they can reach similar inner radii. Indeed in the soft, low-magnetization state the instabilities that can occur, such as the magneto-rotational instability (MRI; \citealt{Balbus91}) and the Rossby Wave Instability (RWI; \citealt{Tagger06}), both tend to heat up the disk more than the instabilities known to occur in a fully magnetized disk.
Using the position of the inner-edge of the disk, the four-state classification separates the softer states further. While the MRI can occur independently of the disk position, the RWI dominates when the inner edge of the disk gets closer to the last stable orbit. This will cause further heating of the inner region of the disk while keeping a relatively stable inner edge. As a consequence of the low magnetization, the softer states will tend to have a more limited coronal emission than the harder states. Using a similar approach as in \citet{Varniere02}, we can also infer that the coronal emission will be higher when the RWI is active in the disk; i.e., when the inner edge of the disk is smaller. 

Within this framework, the harder states tend to be explained by having a fully magnetized disk, hence the Accretion-Ejection Instability \citep{Tagger99} causes the transport of angular momentum and the LFQPO. In this scenario \citep{Tagger04,Varniere07} the inner edge of the disk is pushed out (as in a MAD) because of the magnetic flux stored inside the inner region. As the stored magnetic flux gets destroyed due to the magnetic flux being transported in by the accretion flow in the disk, the inner edge of the disk moves towards $R_{\rm ISCO}$. This process, depending on the magnetization of the disk, is slow enough for the state to be ``steady'' during one observation window. This process is also at the origin of the radio emission visible in those states. 
Once again, depending on whether the inner edge of the disk stays close to the last stable orbit, the RWI will simultaneously occur in the inner region but this time, as shown in \citet{Varniere02}, the coronal emission will be higher when the disk is away from its last stable orbit. 

While a one-on-one association with the different sub-branches of steady-hard and steady soft is difficult, the level of magnetization in the disk may account for the behavioral differences we have observed. 


\subsection{The X-ray--radio correlations of the G12-like and R10 observations within the steady-hard}
\label{sec:x-ray-radio}

We find evidence of a (non-linear) X-ray--radio correlation for GRS1915 that occurs during the decline into two luminosity dips in the steady-hard light curves (Figure~\ref{fig:logflux_lc}). When examining the coronal parameters of these G12-like points, which are limited to the decline period, we find changes which suggest a move towards more canonical hard state characteristics (Section~\ref{sec:corona}). The changes are as follows: (1) While all steady-hard observations show significantly low values of \Efold, these observations show an increase in \Efold\ which corresponds to a decrease in the curvature of the Compton component. This shows a move towards a less curved power law, typical of hard state BHBs. (2) A majority of the steady-hard observations show $1.4 < \Gamma < 2.1$ with a mean value $1.69 \pm 0.26$, which is typical for a hard state. These specific observations occupy a compact spot in the middle of this range and show a much less variability ($\Gamma = 1.72 \pm 0.09$) suggesting a separate population with a more regular power-law index. 

Our fit to the G12-like points matches the slope of the upper G12~track (\citealt{Gallo12}, see also \citealt{Gallo03, Gallo06, Corbel08}) although we note that our uncertainty is large. The changes observed in the coronal parameters within the G12 points increase our confidence that they might map to one of the G12~tracks which are defined using canonical hard state BHBs. Finally, the G12-points lie almost exactly on the upper G12~track (red line in Figure~\ref{fig:logflux}) leading us to believe that these might be one and the same.

Our fit to the R10 points obtained using isolated coronal flux (i.e. without the contribution of disk component) produces a log slope \plateauslope, which is much lower than that observed by \citet{Rushton10}. We find that our value for the slope is consistent with the lower G12~track (blue line in Figure~\ref{fig:logflux}). The two slopes observed within the steady-hard observations of GRS1915 is reminiscent of the two slopes observed in the hard state observations of H1743-322 \citep{Coriat11}.

\subsection{Effects due to high spin in the system}

One of the most noticeable differences in GRS1915 when compared to canonical BHBs is the unusual curvature observed in the coronal component. Our analyses quantify this as \Efold\ values of $\sim$10 keV for the steady-soft observations and 10-30 keV for the majority of the steady-hard observations (see Figure~\ref{fig:corona}c). This curvature has been a major source of difficulty in modeling the spectra of this system. We explained earlier using the classical interpretation for Comptonization, that this curvature can be a result of a corona which is intensely cooled by the ample availability of soft photons from the hot inner disk. This leads us to the obvious next question: what causes the disk in GRS1915 to be so hot? The explanation seems to be high spin ($a_{*} \ge 0.98$; \citealt{McClintock06b}). The extreme spin, which allows the spacial dimension of $R_{\rm ISCO}$ to shrink down to limits smaller than for other known BHBs, could result in the production of a vast number of soft photons that can cool the corona intensely. This will lower the energy of the coronal electrons inhibiting the upscatter the soft photons to high enough energies to produce the usual power law observed in canonical BHBs.

\section{Summary}

\begin{enumerate}
\item  The steady state observations of GRS1915 naturally separate into two groups on the CD with a gap in the middle. We label these two groups, steady-soft and steady-hard.

\item The spectral continuum of GRS1915 displays significant curvature of the Compton component (\Efold$\sim$10 keV for steady-soft and 10-30 keV for a majority of steady-hard), rendering commonly used models such as {\sc powerlaw} or {\sc simpl} ineffective in representing the Comptonization. The new cut-off power-law model {\sc simplcut}, describes the curvature well. The extreme curvature suggests significant cooling of the corona in GRS1915.

\item The disk in GRS1915 is significantly hotter than canonical BHB disks. The temperatures of the disk in a majority of the steady-soft and steady-hard observations are $\sim$2 keV and $\sim$1.2 keV, respectively. Other BHBs show soft and hard observations with typical temperatures of $\le1$ keV and $\le0.5$ keV, respectively.

\item 96\% of the steady-soft observations display an almost constant \Rin\ which is usually observed in canonical soft-state observations of BHBs. However, 4 \% of the steady-soft observations show an increase in \Rin\ at low X-ray luminosity.

\item The steady-hard observations exhibit the presence of a truncated disk with varying \Rin.  The minimum \Rin\ in the steady-hard observations agrees well with the \Rin\ values for the steady-soft observations (and is therefore close to $R_{\rm ISCO}$), but increases by a factor of 4 at highest luminosity. 

\item A strong correlation is noted between disk and coronal flux in the steady-hard observations. The steady-soft observations show only a very weak correlation with comparatively less coronal flux, suggesting that the two coronae have different origin.

\item A sub-population within the steady-hard observations follows the upper G12~track during the luminosity decline period of two dips in the X-ray light curve ($50420<$MJD$<50550$ and $53820<$MJD$<53850$). GRS1915 shows a shift toward a canonical hard state when dropping down to these low luminosities.

\item The R10 observations display a slope which is consistent with the lower G12~track (0.98).

\item It has been shown earlier that the steady observations of GRS1915 yield spectra that resemble the thermal and hard states. We find evidence to suggest that all three states (thermal, SPL and hard) exists within the steady observations of GRS1915.

\item The locations of the thermal and SPL-qpo observations appear to correspond to the location of variable state B in the CD, while the location of the SPL-noqpo observations and the steady-hard observations may correspond to the locations of A and C, respectively \citep{Belloni00}. This suggests a link in the phenomenology of GRS1915 between short and long timescales.
 
\end{enumerate}

\section*{Acknowledgments}

We acknowledge the use of the {it Smithsonian Institution High Performance Cluster (SI/HPC)} for spectral fitting. We would also like to thank Jeff McClintock, Ramesh Narayan, Javier Garcia and Malgosia Sobolewska for interesting and helpful discussions on the content of this paper. JFS has been supported by NASA Einstein Fellowship grant PF5-160144. SDV acknowledges partial support through a Smithsonian Institution CGPS grant. JR acknowledges financial support from the French Research National Agency: CHAOS project ANR-12-BS05-0009 (\texttt{http://www.chaos-project.fr}). PV and JR acknowledge financial support from the UnivEarthS Labex program at Sorbonne Paris Cit\'e (ANR-10-LABX-0023 and ANR-11-IDEX-0005-02).

\bibliography{grs1915,biblio}

\newpage

\begin{center}
\begin{longtable}{@{}lcccccc}
\caption[Steady-soft and steady-hard Intervals of GRS1915+105]{Observation list} \label{obslist} \\


\hline \\ Reference$^1$ & MJD$^2$ & MET$^3$ Start & MET$^3$ Stop & Exposure (s) & Type & Spectral fit$^4$ \\ \\ \hline \\
\endfirsthead

{{\tablename\ \thetable{} -- continued }} \\ \\

\hline \\ Reference$^1$ & MJD$^2$ & MET$^3$ Start & MET$^3$ Stop & Exposure (s) & Type & Spectral fit$^4$ \\ \\ \hline \\
\endhead

\\ \hline \\ Continued on next page \\ 
\endfoot

\hline \hline
\endlastfoot

0829\textunderscore a  & 50182.71652 & 71686083 & 71688931 & 2848 & soft & yes \\
0829\textunderscore b  & 50182.78550 & 71691827 & 71695107 & 3280 & soft & yes \\
0829\textunderscore c  & 50182.85309 & 71697587 & 71701027 & 3440 & soft & yes \\
0837a\textunderscore a  & 50190.58735 & 72366723 & 72368371 & 1648 & soft & no \\
0837a\textunderscore b  & 50190.65800 & 72372739 & 72374563 & 1824 & soft & yes \\
0837b\textunderscore a  & 50190.71884 & 72378019 & 72379795 & 1776 & soft & no \\
0837c\textunderscore a  & 50190.77772 & 72383379 & 72384611 & 1232 & soft & no \\
0840\textunderscore a  & 50193.44735 & 72613923 & 72615379 & 1456 & soft & no \\
\end{longtable}
\begin{table}[h]
\begin{tabular}{lc}

$^1$ Data reference name, using MET day (see note 3). \\
$^2$ Modified Julian Date of exposure midpoint; MDJ = JD - 2400000.5. \\
$^3$ Mission Elapsed Time (MET), here in integer seconds, is the time since 1994 January 1 0 h (UTC). The truncated fraction is 0.374 s. \\ RXTE Standard 2 data is collected during 16-s intervals of Spacecraft time (SCT), which is related by: MET = SCT + 3.374 \\
$^4$ Indicates whether a good-fit (i.e. a fit with $\chi^{2}_{\nu} <$ 2) was achieved or not. \\

\end{tabular}
\end{table}
\end{center}

\end{document}